\documentclass[11pt, a4paper, oneside]{article}
\usepackage{amsmath}
\usepackage{amssymb}
\usepackage{amsfonts}

\begin{document}

\newpage 
\begin{titlepage}
\begin{flushright}
UFSCARF-TH-09-11
\end{flushright}
\vskip 3.0cm
\begin{center}
{\Large  Reflection matrices for the $U_{q}[osp(r|2m)^{(1)}]$ vertex model }\\
\vskip 2cm
{\large A. Lima-Santos} \\
\vskip 1cm
{\em Universidade Federal de S\~ao Carlos\\
Departamento de F\'{\i}sica \\
C.P. 676, 13565-905, S\~ao Carlos-SP, Brasil}\\
\end{center}
\vskip 2cm

\begin{abstract}
The graded reflection equation is investigated for the $U_{q}[osp(r|2m)^{(1)}]$ vertex model. We have found four classes of diagonal
solutions with at the most one free parameter and twelve classes of non-diagonal ones with the number of free parameters depending on
the number of bosonic ($r$) and fermionic ($2m$) degrees of freedom.
\end{abstract}

\vskip 2.5cm
\centerline{{\small PACS numbers:  05.50+q, 02.30.IK, 75.10.Jm}}
\vskip 0.1cm
\centerline{{\small Keywords: Reflection Equations, K-matrices, Superalgebras}}
\vskip 2.5cm
\centerline{{\today}}
\end{titlepage}

\section{Introduction}

Integrability in classical vertex models and quantum spin chains is
intimately connected with solutions of the Yang-Baxter equation \cite{BAX}.
This equation plays a central role in the Quantum Inverse Scattering Method
which provides an unified approach to construct and study physical
properties of integrable models \cite{QISM,KOR}. Usually these systems are
studied with periodic boundary conditions but more general boundaries can
also be included in this framework as well. Physical properties associated
with the bulk of the system are not expected to be influenced by boundary
conditions in the thermodynamical limit. Nevertheless, there are surface
properties such as the interfacial tension where the boundary conditions are
of relevance. Moreover, the conformal spectra of lattice models at
criticality can be modified by the effect of boundaries \cite{CAR1}.

Integrable systems with open boundary conditions can also be accommodated
within the framework of the Quantum Inverse Scattering Method \cite{SK}. In
addition to the solution of the Yang-Baxter equation governing the dynamics
of the bulk there is another fundamental ingredient, the reflection matrices 
\cite{CHER}. These matrices, also referred as $K$-matrices, represent the
interactions at the boundaries and compatibility with the bulk integrability
requires these matrices to satisfy the so-called reflection equations \cite%
{SK,CHER}.

At the moment, the study of general regular solutions of the reflection
equations for vertex models based on $q$-deformed Lie algebras \cite{BAZ,JIM}
has been successfully accomplished. See \cite{LIM} for instance and
references therein. However, this same analysis for vertex models based on
Lie superalgebras are still restricted to diagonal solutions associated with
the $U_{q}[sl(m|n)]$ \cite{G1,BRA} and $U_{q}[osp(2|2)]$ symmetries \cite%
{GUA} and non-diagonal solutions related to super-Yangians $osp(m|n)$ \cite%
{G2} and $sl(m|n)$ \cite{G3,GALLEAS5}.

The aim of this paper is to touch again in the classification of the
solutions of the reflection equations based in the Lie superalgebras already
initialized in \cite{LIMG} with the $U_{q}[sl(r|2m)^{(2)}]$ vertex model. \
Here we will present the most general set of solutions of the reflection
equation for the $U_{q}[osp(r|2m)^{(1)}]$ vertex model, keeping the
structure presented in\cite{LIMG}. This paper is organized as follows. In
the next section we present the $U_{q}[osp(r|2m)^{(1)}]$ vertex model. This
supplies the way for the analysis of the corresponding reflection equations
and in the section $3$ we present four classes of diagonal solutions. In the
section $4$ we present twelve classes of non-diagonal solutions what we hope
to be the most general set of $K$-matrices for the vertex model here
considered. Concluding remarks are discussed in the section $5$, and in the
appendices $A$ and $B$ we present special solutions associated with the $%
U_{q}[osp(1|2)^{(1)}]$ and $U_{q}[osp(2|2)^{(1)}]$ cases respectively.
Finally, in the appendix $C$ we describe the main steps of our construction.

\section{The $U_{q}[osp(r|2m)^{(1)}]$ vertex model}

Classical vertex models of statistical mechanics are nowadays well known to
play a fundamental role in the theory of two-dimensional integrable systems 
\cite{BAX}. In this sense, it turns out that a $R$-matrix satisfying the
Yang-Baxter equation gives rise to the Boltzmann weights of an exactly
solvable vertex model. The Yang-Baxter equation consist of an operator
relation for a complex valued matrix $R:\mathbb{C}\rightarrow \mathrm{End}$ $%
\left( V\otimes V\right) $ reading 
\begin{equation}
R_{12}(x)R_{13}(xy)R_{23}(y)=R_{23}(y)R_{13}(xy)R_{12}(x),  \label{YB}
\end{equation}%
where $R_{ij}(x)$ refers to the $R$-matrix acting non-trivially in the $i$th
and $j$th spaces of the tensor product $V\otimes V\otimes V$ and the complex
variable $x$ denotes the spectral parameter. Here $V$ is a finite
dimensional $Z_{2}$ graded linear space and the tensor products appearing in
the above definitions should be understood in the graded sense. For
instance, we have $\left[ A\otimes B\right] _{j%
\;k}^{il}=A_{j}^{i}B_{k}^{l}(-1)^{(p_{i}+p_{j})p_{l}}$ for generic matrices $%
A$ and $B$. The Grassmann parities $p_{i}$ assume values on the group $Z_{2}$
and enable us to distinguish bosonic and fermionic degrees of freedom. More
specifically, the $\alpha $th degree of freedom is distinguished by the
Grassmann parity $p_{i}=0(1)$ for $i$ bosonic (fermionic).

An important class of solutions of the Yang-Baxter equation (\ref{YB}) is
denominated trigonometric $R$-matrices containing an additional parameter $q$
besides the spectral parameter. Usually such $R$-matrices have their roots
in the $U_{q}[\mathcal{G}]$ quantum group framework, which permit us to
associate a fundamental trigonometric $R$-matrix to each Lie algebra or Lie
superalgebra $\mathcal{G}$ \cite{BAZ,JIM,SHA}. In particular, explicit $R$%
-matrices were exhibited in \cite{GALLEAS1,GALLEAS2} for a variety of
quantum superalgebras in terms of standard Weyl matrices, providing in this
way a suitable basis for the analysis of the corresponding reflection
equation.

\noindent The $U_{q}[osp(r|2m)^{(1)}]$ invariant $R$-matrices are given by 
\begin{eqnarray}
R(x) &=&\sum_{\overset{i=1}{i\neq i^{\prime }}}^{N}(-1)^{p_{i}}a_{i}(x)\hat{e%
}_{ii}\otimes \hat{e}_{ii}+b(x)\sum_{\overset{i,j=1}{i\neq j,i\neq j^{\prime
}}}^{N}\hat{e}_{ii}\otimes \hat{e}_{jj}  \notag \\
&&+{\bar{c}}(x)\sum_{\overset{i,j=1}{i<j,i\neq j^{\prime }}}(-1)^{p_{i}p_{j}}%
\hat{e}_{ji}\otimes \hat{e}_{ij}+c(x)\sum_{\overset{i,j=1}{i>j,i\neq
j^{\prime }}}^{N}(-1)^{p_{i}p_{j}}\hat{e}_{ji}\otimes \hat{e}_{ij}  \notag \\
&&+\sum_{i,j=1}^{N}(-1)^{p_{i}}d_{i,j}(x)\hat{e}_{ij}\otimes \hat{e}%
_{i^{\prime }j^{\prime }}  \label{Rsl}
\end{eqnarray}%
where $N=r+2m$ is the dimension of the graded space with $r$ bosonic and $2m$
fermionic degrees of freedom. Here $i^{\prime }=N+1-i$ corresponds to the
conjugated index of $i$ and $\hat{e}_{ij}$ refers to a usual $N\times N$
Weyl matrix with only one non-null entry with value $1$ at the row $i$ and
column $j$.

In \cite{GALLEAS3} it was demonstrated that the use of an appropriate
grading structure plays a decisive role in the investigation of the
thermodynamic limit and finite size properties of integrable quantum spin
chains based on superalgebras. In what follows we shall adopt the grading
structure 
\begin{equation}
p_{i}=\left\{ 
\begin{array}{c}
0\qquad \mathrm{for\ }i=1,...,m\ \mathrm{\ and\ \ }i=r+m+1,...,N \\ 
\ 1\quad \mathrm{for\ }i=m+1,...,r+m\qquad \qquad \mathrm{\qquad \qquad
\quad }%
\end{array}%
\right. ,  \label{grad}
\end{equation}%
and the corresponding Boltzmann weights $a_{i}(x)$, $b(x)$, $c(x)$, $\bar{c}%
(x)$ and $d_{ij}(x)$ are then given by%
\begin{eqnarray}
a_{i}(x) &=&(x-\zeta )(x^{(1-p_{i})}-q^{2}x^{p_{i}}),\quad \
b(x)=q(x-1)(x-\zeta ),  \notag \\
c(x) &=&(1-q^{2})(x-\zeta ),\qquad \qquad \quad {\bar{c}}(x)=x(1-q^{2})(x-%
\zeta )
\end{eqnarray}%
and 
\begin{equation}
\;d_{i,j}(x)=\left\{ 
\begin{array}{c}
q(x-1)(x-\zeta )+x(q^{2}-1)(\zeta -1),\qquad \quad \quad 
\hfill%
(i=j=j^{\prime }) \\ 
(x-1)[(x-\zeta )(-1)^{p_{i}}q^{2p_{i}}+x(q^{2}-1)],\qquad 
\hfill%
(i=j\neq j^{\prime }) \\ 
(q^{2}-1)[\zeta (x-1)\frac{\theta _{i}q^{t_{i}}}{\theta _{j}q^{t_{j}}}%
-\delta _{i,j^{\prime }}(x-\zeta )],\qquad \quad \quad 
\hfill%
(i<j) \\ 
(q^{2}-1)x[(x-1)\frac{\theta _{i}q^{t_{i}}}{\theta _{j}q^{t_{j}}}-\delta
_{i,j^{\prime }}(x-\zeta )],\qquad \quad \quad 
\hfill%
(i>j)%
\end{array}%
\right.  \label{wb2}
\end{equation}%
where $\zeta =q^{r-2m-2}$. The remaining variables $\theta _{i}$ and $t_{i}$
depend strongly on the grading structure considered and they are determined
by the relations%
\begin{equation}
\theta _{i}=\left\{ 
\begin{array}{c}
(-1)^{-\frac{p_{i}}{2}},\qquad 
\hfill%
1\leq i<\frac{N+1}{2} \\ 
1,%
\hfill%
i=\frac{N+1}{2} \\ 
(-1)^{\frac{p_{i}}{2}},\qquad 
\hfill%
\frac{N+1}{2}<i\leq N%
\end{array}%
\right.
\end{equation}%
\begin{equation}
t_{i}=\left\{ 
\begin{array}{c}
\displaystyle%
i+[\frac{1}{2}-p_{i}+2\sum_{i\leq j<\frac{N+1}{2}}p_{j}],\qquad 
\hfill%
1\leq i<\frac{N+1}{2} \\ 
\displaystyle%
\frac{N+1}{2},%
\hfill%
i=\frac{N+1}{2} \\ 
\displaystyle%
i-[\frac{1}{2}-p_{i}+2\sum_{\frac{N+1}{2}<j\leq i}p_{j}],\qquad 
\hfill%
\frac{N+1}{2}<i\leq N%
\end{array}%
\right.  \label{bwf}
\end{equation}

The $R$-matrix (\ref{Rsl}) satisfies important symmetry relations, besides
the standard properties of regularity and unitarily, namely 
\begin{eqnarray}
&&\mathrm{PT-Symmetry:\;}\;\;\;\;\;\;\;\;\;\;\;\;\;\;%
\;R_{21}(x)=R_{12}^{st_{1}st_{2}}(x)  \notag \\
&&\mathrm{Cros}\text{\textrm{sing}}\mathrm{\;Symmetry:}\;\;\;\;\;\;R_{12}(x)=%
\frac{\rho (x)}{\rho (x^{-1}\eta ^{-1})}V_{1}R_{12}^{st_{2}}(x^{-1}\eta
^{-1})V_{1}^{-1},
\end{eqnarray}%
where the symbol $st_{k}$ stands for the supertransposition operation in the
space with index $k$. In its turn $\rho (x)$ is an appropriate normalization
function given by $\rho (x)=q(x-1)(x-\zeta )$ and the crossing parameter is $%
\eta =\zeta ^{-1}$. At this stage it is convenient to consider the $%
U_{q}[osp(2n|2m)^{(1)}]$ and the $U_{q}[osp(2n+1|2m)^{(1)}]$ vertex models
separately and their corresponding crossing matrix $V$ is an anti-diagonal
matrix with the following non-null entries $V_{i{i}^{^{\prime }}}$,

\begin{itemize}
\item $U_{q}\left[ osp(2n|2m)^{(1)}\right] $: 
\begin{equation}
V_{i{i}^{^{\prime }}}=\left\{ 
\begin{array}{c}
\displaystyle%
(-1)^{\frac{1-p_{i}}{2}}q^{\digamma _{1}^{(i)}},\qquad 
\hfill%
1\leq i\leq \frac{N}{2} \\ 
\displaystyle%
(-1)^{\frac{1+p_{i}}{2}}q^{\digamma _{2}^{(i)}},\qquad 
\hfill%
\frac{N}{2}+1\leq i\leq N%
\end{array}%
\right.
\end{equation}

\item $U_{q}\left[ osp(2n+1|2m)^{(1)}\right] $: 
\begin{equation}
V_{i{i}^{^{\prime }}}=.\left\{ 
\begin{array}{c}
\displaystyle%
(-1)^{\frac{1-p_{i}}{2}}q^{\digamma _{1}^{(i)}},\qquad 
\hfill%
1\leq i\leq \frac{N-1}{2}, \\ 
\displaystyle%
(-1)^{\frac{1-p_{i}}{2}}q^{\digamma _{3}^{(i)}},\qquad 
\hfill%
i=\frac{N+1}{2}, \\ 
\displaystyle%
(-1)^{\frac{1+p_{i}}{2}}q^{\digamma _{4}^{(i)}},\qquad 
\hfill%
\frac{N+3}{2}\leq i\leq N.%
\end{array}%
\right.
\end{equation}%
where%
\begin{eqnarray}
\displaystyle%
\digamma _{1}^{(i)} &=&i-1+p_{1}-p_{i}-2\sum_{j=1}^{i-1}p_{j},\quad 
\displaystyle%
\digamma _{2}^{(i)}=i-2-p_{1}-p_{i}-2\sum_{j=2\neq \frac{N}{2}+1}^{i-1}p_{j},
\notag \\
\displaystyle%
\digamma _{3}^{(i)} &=&\frac{N}{2}-1-p_{1}-p_{i}-2\sum_{j=2}^{\frac{N-1}{2}%
}p_{j},\quad 
\displaystyle%
\digamma _{4}^{(i)}=i-2-p_{1}-p_{i}-2\sum_{j=2}^{i-1}p_{j}.
\end{eqnarray}
\end{itemize}

The construction of integrable models with open boundaries was largely
impulsed by Sklyanin's pioneer work \cite{SK}. In Sklyanin's approach the
construction of such models are based on solutions of the so-called
reflection equations \cite{CHER,SK} for a given integrable bulk system. The
reflection equations determine the boundary conditions compatible with the
bulk integrability and it reads 
\begin{equation}
R_{21}(x/y)K_{2}^{-}(x)R_{12}(xy)K_{1}^{-}(y)=K_{1}^{-}(y)R_{21}(xy)K_{2}^{-}(x)R_{12}(x/y),
\label{RE}
\end{equation}%
where the tensor products appearing in (\ref{RE}) should be understood in
the graded sense. The matrix $K^{-}(x)$ describes the reflection at one of
the ends of an open chain while a similar equation should also hold for a
matrix $K^{+}(x)$ describing the reflection at the opposite boundary. As
discussed above, the $U_{q}[osp(r|2m)^{(1)}]$ $R$-matrix satisfies important
symmetry relations such as the PT-symmetry and crossing symmetry. When these
properties are fulfilled one can follow the scheme devised in \cite{BRA,MEZ}
and the matrix $K^{-}(x)$ is obtained by solving the Eq. (\ref{RE}) while
the matrix $K^{+}(x)$ can be obtained from the isomorphism $K^{-}(x)\mapsto
K^{+}(x)^{st}=K^{-}(x^{-1}\eta ^{-1})V^{st}V$.

The purpose of this work is to investigate the general families of regular
solutions of the graded reflection equation (\ref{RE}). Regular solutions
mean that the $K$-matrices have the general form 
\begin{equation}
K^{-}(x)=\sum_{i,j=1}^{N}k_{i,j}(x)\;\hat{e}_{ij},  \label{KM}
\end{equation}%
such that the condition $k_{i,j}(1)=\delta _{ij}$ holds for all matrix
elements.

The direct substitution of (\ref{KM}) and the $U_{q}[osp(r|2m)^{(1)}]$ $R$%
-matrix (\ref{Rsl})-(\ref{bwf}) in the graded reflection equation (\ref{RE}%
), leave us with a system of $N^{4}$ functional equations for the entries $%
k_{i,j}(x)$. In order to solve these equations we shall make use of the
derivative method. Thus, by differentiating the equation (\ref{RE}) with
respect to $y$ and setting $y=1$, we obtain a set of algebraic equations for
the matrix elements $k_{i,j}(x)$. Although we obtain a large number of
equations only a few of them are actually independent and a direct
inspection of those equations, in the lines described in \cite{LIM} and \cite%
{LIMG}, allows us to find the branches of regular solutions. In what follows
we shall present our findings for the regular solutions of the reflection
equation associated with the $U_{q}[osp(r|2m)^{(1)}]$ vertex model. We have
obtained four families of diagonal solutions and twelve families of
non-diagonal ones. The special solutions associated with the cases $%
U_{q}[osp(1|2)^{(1)}]$ and $U_{q}[osp(2|2)^{(1)}]$ are presented in the
appendices A and B respectively.

\section{Diagonal K-matrix solutions}

The diagonal solutions of the graded reflection equation (\ref{RE}) is
characterized by a $K$-matrix of the form 
\begin{equation}
K^{-}(x)=\sum_{i=1}^{N}k_{i,i}(x)\hat{e}_{ii}.  \label{km}
\end{equation}%
with the entries $k_{i,j}(x)$ related with $k_{1,1}(x)$ in a general form
given by%
\begin{equation}
k_{i,i}(x)=\frac{(\beta _{i,i}-\beta _{1,1})(x-1)+2}{(\beta _{i,i}-\beta
_{1,1})(x^{-1}-1)+2}k_{1,1}(x)
\end{equation}%
for $i=2,...,N-1$ and%
\begin{equation}
k_{N,N}(x)=\frac{(\beta _{N,N}-\beta _{N-1,N-1})(x-1)+2}{(\beta _{N,N}-\beta
_{N-1,N-1})(x^{-1}-1)+2}k_{N-1,N-1}(x)
\end{equation}%
The parameters $\beta _{i,i}=\frac{d}{dx}[k_{i,i}(x)]_{x=1}$ are constrained
by the reflection equations and forced to fall in the four families of
diagonal $K$-matrices that we shall refer as solutions of type $\mathcal{D}%
_{1}$ to type $\mathcal{D}_{4}$.

\begin{itemize}
\item \textbf{Solution $\mathcal{D}_{1}$:} \ Family formed by solutions
without free parameters characterized by the label $p$ assuming discrete
values in the interval $2\leq p\leq m+1$.%
\begin{eqnarray}
k_{1,1}(x) &=&\cdots =k_{p-1,p-1}(x)=1,  \notag \\
k_{p,p}(x) &=&\cdots =k_{N+1-p,N+1-p}(x)=\frac{x+\epsilon q^{2p-1}\sqrt{%
\zeta }}{x^{-1}+\epsilon q^{2p-1}\sqrt{\zeta }},  \notag \\
k_{N+2-p,N+2-p}(x) &=&\cdots =k_{N,N}(x)=x^{2}.
\end{eqnarray}%
Here and in what follows, $\epsilon $ is a discrete parameter assuming the
values $\pm 1$.

\item \textbf{Solution $\mathcal{D}_{2}$:} Family formed by solutions
without free parameters. The discrete label $p$ can assume values in the
interval $m+2\leq p<\frac{N+1}{2}$%
\begin{eqnarray}
k_{1,1}(x) &=&\cdots =k_{p-1,p-1}(x)=1,  \notag \\
k_{p,p}(x) &=&\cdots =k_{N+1-p,N+1-p}(x)=\frac{x+\epsilon q^{4m+3-2p}\sqrt{%
\zeta }}{x^{-1}+\epsilon q^{4m+3-2p}\sqrt{\zeta }},  \notag \\
k_{N+2-p,N+2-p}(x) &=&\cdots =k_{N,N}(x)=x^{2}.
\end{eqnarray}%
In addition to these free parameter solutions we have more two families of
one-parameter solutions valid only for the models with $r$ even,

\item \textbf{Solution $\mathcal{D}_{3}$:} Class of solution valid only for $%
r=2n$ with $n\geq 1$%
\begin{eqnarray}
k_{1,1}(x) &=&\cdots =k_{n+m-1,n+m-1}(x)=1,  \notag \\
k_{n+m,m+m}(x) &=&\frac{\beta (x-1)+2}{\beta (x^{-1}-1)+2},  \notag \\
k_{n+m+1,n+m+1}(x) &=&\frac{\beta (x-q^{2}\zeta ^{-1})-2x}{\beta
(x^{-1}-q^{2}\zeta ^{-1})-2x^{-1}},  \notag \\
k_{n+m+2,n+m+2}(x) &=&\cdots =k_{N,N}(x)=x^{2}.
\end{eqnarray}

\item \textbf{Solution $\mathcal{D}_{4}$:} Class of solution valid only for $%
r=2n$ with $n\geq 1$%
\begin{eqnarray}
k_{1,1}(x) &=&\cdots =k_{n+m-1,n+m-1}(x)=1,  \notag \\
k_{n+m,m+m}(x) &=&\cdots =k_{N,N}(x)=\frac{\beta (x-1)+2}{\beta (x^{-1}-1)+2}%
.
\end{eqnarray}
\end{itemize}

In the families \textbf{$\mathcal{D}_{3}$} and \textbf{$\mathcal{D}_{4}$}
the free parameter is defined by $\beta =\beta _{n+m,n+m}-\beta _{1,1}.$

Thus, for the $osp(2n+1|2m)^{(1)}$ we have $n+m$ diagonal solutions without
free parameters and for the $osp(2n|2m)^{(1)}$ we have $n+m+1$ diagonal
solutions from which two solutions have one free parameter.

\section{Non-Diagonal K-matrix Solutions}

\bigskip Analyzing the reflection matrix equation (\ref{RE}) we can see that
the non-diagonal elements $k_{i,j}(x)$ with fermionic labels\ ( $i\neq
j=1,..,m$ and $i\neq j=m+r+1,...,N$) are write in terms of $k_{1,N}(x)$ 
\begin{equation}
k_{i,j}(x)=\left\{ 
\begin{array}{c}
\mathcal{F}(x)\left( \beta _{i,j}c_{1}(x)d_{1,1}(x)+\beta _{j^{\prime
},i^{\prime }}b(x)d_{i,j^{\prime }}(x)\right) ,\quad j<i^{\prime } \\ 
\beta _{i,i^{\prime }}\frac{k_{1,N}(x)}{\beta _{1,N}},%
\hfill%
i=j^{\prime } \\ 
\mathcal{F}(x)\left( \beta _{i,j}c_{2}(x)d_{1,1}(x)+\beta _{j^{\prime
},i^{\prime }}b(x)d_{i,j^{\prime }}(x)\right) ,\quad j>i^{\prime }%
\end{array}%
\right.
\end{equation}%
where%
\begin{equation}
\mathcal{F}(x)=\frac{b^{2}(x)-a_{1}(x)d_{1,1}(x)}{%
b^{2}(x)d_{1,2}(x)d_{2,1}(x)-c_{1}(x)c_{2}(x)d_{1,1}^{2}(x)}\frac{k_{1,N}(x)%
}{\beta _{1,N}}
\end{equation}%
and those matrix elements with the bosonic labels $i\neq j=m+1,...,m+r$ are
write in terms of $k_{m+1,N-m}(x)$ 
\begin{equation}
k_{i,j}(x)=\left\{ 
\begin{array}{c}
\mathcal{G}(x)\left( \beta _{i,j}c_{1}(x)d_{m+1,m+1}(x)+\beta _{j^{\prime
},i^{\prime }}b(x)d_{i,j^{\prime }}(x)\right) ,\quad j<i^{\prime } \\ 
\beta _{i,i^{\prime }}\frac{k_{m+1,N-m}(x)}{\beta _{m+1,N-m}},%
\hfill%
i=j^{\prime } \\ 
\mathcal{G}(x)\left( \beta _{i,j}c_{2}(x)d_{m+1,m+1}(x)+\beta _{j^{\prime
},i^{\prime }}b(x)d_{i,j^{\prime }}(x)\right) ,\quad j>i^{\prime }%
\end{array}%
\right.
\end{equation}%
where%
\begin{equation}
\mathcal{G}(x)=\frac{b^{2}(x)-a_{m+1}(x)d_{m+1,m+1}(x)}{%
b^{2}(x)d_{1,2}(x)d_{2,1}(x)-c_{1}(x)c_{2}(x)d_{m+1,m+1}^{2}(x)}\frac{%
k_{m+1,N-m}(x)}{\beta _{m+1,N-m}}.
\end{equation}

In general, the non-diagonal elements $k_{i,i}(x)$ have the structure 
\begin{equation}
k_{i,j}(x)=\left\{ 
\begin{array}{c}
\beta _{i,j}xG(x),\quad 
\hfill%
\mathrm{\ }i>j^{\prime } \\ 
\beta _{i,j}xG(x),\quad 
\hfill%
i>j^{\prime } \\ 
\beta _{i,j}G(x)H_{f}(x),\quad 
\hfill%
\mathrm{\quad }i=j^{\prime }(\ \mathrm{fermionic}) \\ 
\beta _{i,j}G(x)H_{b}(x),\quad 
\hfill%
\mathrm{\quad }i=j^{\prime }(\ \mathrm{bosonic})%
\end{array}%
\right.  \label{Gf}
\end{equation}%
Here and in what follows $G(x)$ is an arbitrary function satisfying the
regular condition $k_{i,j}(1)=\delta _{ij}$ and%
\begin{equation}
\beta _{i,j}=\frac{d}{dx}[k_{i,j}(x)]_{x=1},\qquad H_{f}(x)=\frac{x-\epsilon
q\sqrt{\zeta }}{1-\epsilon q\sqrt{\zeta }},\qquad H_{b}(x)=\frac{qx+\epsilon 
\sqrt{\zeta }}{q+\epsilon \sqrt{\zeta }}.  \label{Hs}
\end{equation}%
Moreover, the diagonal entries $k_{i,i}(x)$ satisfy defined recurrence
relations which depend on the bosonic and fermionic degree of freedom.
However, for the $U_{q}[osp(r|2m)^{(1)}]$ \ model we have found $K$-matrix
solutions with both degree of freedom only for the cases with $m=1$ and for
the cases with $r=1$ and $r=2$.

In this section we shall focus on the non-diagonal solutions of the graded
reflection equation (\ref{RE}). We have found twelve classes of non-diagonal
solutions that we refer in what follows as solutions of type $\mathcal{M}%
_{1} $ to type $\mathcal{M}_{12}$. Explicitly, we have three classes of
solutions: solutions with only fermionic degree of freedom, named fermionic $%
K$-matrices; five classes of solutions with only bosonic degree freedom,
named bosonic $K$-matrices and four solutions with both degree of freedom,
named complete $K$-matrices.

\subsection{Fermionic K-matrices}

Here we shall focus on the non-diagonal solutions of the graded reflection
equation (\ref{RE}) turning off the bosonic degree of freedom \ \textit{i.e.}
$k_{i,j}(x)=0$ \ for $i\neq j=m+1,...,m+r$. We have found three classes of
non-diagonal solutions that we refer in what follows as solutions of type $%
\mathcal{M}_{1}$ to type $\mathcal{M}_{3}$:

\subsubsection{\textbf{Solution $\mathcal{M}_{1}$}}

The solution of type $\mathcal{M}_{1}$ is valid only for the $%
U_{q}[osp(r|2)^{(1)}]$ models with $r\geq 1$ and the $K$-matrix has the
following block structure 
\begin{equation}
K^{-}(x)=\left( 
\begin{array}{ccc}
k_{1,1}(x) & \mathbb{O}_{1\times r} & k_{1,N}(x) \\ 
\mathbb{O}_{r\times 1} & \mathbb{K}_{1}(x) & \mathbb{O}_{r\times 1} \\ 
k_{N,1}(x) & \mathbb{O}_{1\times r} & k_{N,N}(x)%
\end{array}%
\right) ,
\end{equation}%
where $\mathbb{O}_{a\times b}$ is a $a\times b$ null matrix and 
\begin{equation}
\mathbb{K}_{1}(x)=\left[ x-\frac{1}{2}\beta (x-1)(q^{2}\zeta -x)\right] 
\mathbb{I}_{r\times r}.
\end{equation}%
Here and in what follows $\mathbb{I}_{r\times r}$ denotes a $r\times r$
identity matrix and the remaining non-null entries are given by 
\begin{eqnarray}
k_{1,1}(x) &=&x-\frac{1}{2}\beta (x-1)(q^{2}\zeta +1),  \notag \\
k_{1,N}(x) &=&\frac{1}{2}\beta _{1,N}(x^{2}-1),\quad k_{N,1}(x)=-\frac{1}{2}%
\frac{\beta ^{2}}{\beta _{1,N}}q^{2}\zeta (x^{2}-1),\;  \notag \\
k_{N,N}(x) &=&x-\frac{x}{2}\beta (x-1)(q^{2}\zeta +1).
\end{eqnarray}%
where $\beta =\beta _{2,2}-\beta _{1,1}$ and $\beta _{1,N}$ are two free
parameters. We remark here that this solution for $r=2$ consist of a
particular case of the three parameter solution given in the appendix $B$
for the $U_{q}[osp(2|2)^{(1)}]$ vertex model.

\subsubsection{\textbf{Solution $\mathcal{M}_{2}$}}

The $U_{q}[osp(r|4)^{(1)}]$ vertex models admit the solution $\mathcal{M}%
_{2} $ whose corresponding $K$-matrix has the following structure 
\begin{equation}
K^{-}(x)=\left( 
\begin{array}{ccccc}
k_{1,1}(x) & k_{1,2}(x) &  & k_{1,N-1}(x) & k_{1,N}(x) \\ 
k_{2,1}(x) & k_{2,2}(x) & \mathbb{O}_{2\times r} & k_{2,N-1}(x) & k_{2,N}(x)
\\ 
& \mathbb{O}_{r\times 2} & \mathbb{K}_{2}(x) & \mathbb{O}_{r\times 2} &  \\ 
k_{N-1,1}(x) & k_{N-1,2}(x) & \mathbb{O}_{2\times r} & k_{N-1,N-1}(x) & 
k_{N-1,N}(x) \\ 
k_{N,1}(x) & k_{N,2}(x) &  & k_{N,N-1}(x) & k_{N,N}(x)%
\end{array}%
\right) ,
\end{equation}%
where $\mathbb{K}_{2}(x)=k_{3,3}(x)\mathbb{I}_{r\times r}$. The non-diagonal
entries can be written as 
\begin{eqnarray}
k_{1,2}(x) &=&\beta
_{1,2}G(x)\;\;\;\;\;\;\;\;\;\;\;\;\;\;\;\;\;\;\;\;\;\;\;\;\;\;\;\;\;\;\;%
\;k_{2,1}(x)=\beta _{2,1}G(x)  \notag \\
k_{1,N-1}(x) &=&\beta
_{1,N-1}G(x)\;\;\;\;\;\;\;\;\;\;\;\;\;\;\;\;\;\;\;\;\;\;\;\;\;\;%
\;k_{N-1,1}(x)=\frac{\theta _{N-1}q^{t_{N-1}}}{\theta _{2}q^{t_{2}}}\frac{%
\beta _{1,2}\beta _{2,1}}{\beta _{1,N-1}}G(x)  \notag \\
k_{2,N-1}(x) &=&-\frac{\beta _{2,1}\beta _{1,N}}{\beta _{1,2}}%
G(x)H_{f}(x)\;\;\;\;\;\;\;\;\;\;\;\;\;\;k_{N-1,2}(x)=q^{r-2}\frac{\beta
_{1,2}\beta _{2,1}\beta _{1,N}}{\beta _{1,N-1}^{2}}G(x)H_{f}(x)  \notag \\
k_{2,N}(x) &=&-\frac{\theta _{2}q^{t_{2}}}{\theta _{1}q^{t_{1}}}\frac{%
\epsilon }{\sqrt{\zeta }}\beta
_{1,N-1}xG(x)\;\;\;\;\;\;\;\;\;\;\;\;\;\;\;\;\;\;\;\;\;\;\;k_{N,2}(x)=-\frac{%
\theta _{N}q^{t_{N}}}{\theta _{2}q^{t_{2}}}\frac{\epsilon }{\sqrt{\zeta }}%
\frac{\beta _{2,1}\beta _{1,2}}{\beta _{1,N-1}}xG(x)  \notag \\
k_{N-1,N}(x) &=&-\frac{\theta _{N-1}q^{t_{N-1}}}{\theta _{1}q^{t_{1}}}\frac{%
\epsilon }{\sqrt{\zeta }}\beta
_{1,2}xG(x)\;\;\;\;\;\;\;\;\;\;\;\;\;\;\;\;k_{N,N-1}(x)=-\frac{\theta
_{N}q^{t_{N}}}{\theta _{2}q^{t_{2}}}\frac{\epsilon }{\sqrt{\zeta }}\beta
_{2,1}xG(x)  \notag \\
k_{N,1}(x) &=&\frac{\theta _{N-1}q^{t_{N-1}}}{\theta _{2}q^{t_{2}}}\frac{%
\beta _{2,1}^{2}\beta _{1,N}}{\beta _{1,N-1}^{2}}G(x)H_{f}(x)\;\;\;\;\;\;\;%
\;\;\;\;\;\;k_{1,N}(x)=\beta _{1,N}G(x)H_{f}(x),
\end{eqnarray}

With respect to the diagonal matrix elements, we have the following
expressions 
\begin{eqnarray}
k_{1,1}(x) &=&[(\beta _{N,N}-\beta _{3,3})x-(\beta _{N,N}-\beta
_{1,1}-2)xH_{f}(x)+\beta _{3,3}-\beta _{1,1}]\frac{G(x)}{x^{2}-1}  \notag \\
&&+[1+\epsilon q^{r-1}\sqrt{\zeta }]\frac{\Delta (x)}{x^{2}-1}
\end{eqnarray}%
for the recurrence relation%
\begin{eqnarray}
k_{2,2}(x) &=&k_{1,1}(x)+(\beta _{2,2}-\beta _{1,1})G(x),  \notag \\
k_{3,3}(x) &=&k_{1,1}(x)+(\beta _{3,3}-\beta _{1,1})G(x)+\Delta (x),  \notag
\\
k_{N-1,N-1}(x) &=&k_{3,3}(x)+\left( \beta _{N-1,N-1}-\beta _{3,3}\right)
xG_{2}(x)+\epsilon q^{r-1}\sqrt{\zeta }\Delta (x),  \notag \\
k_{N,N}(x) &=&x^{2}k_{1,1}(x)+\left( \beta _{N,N}-\beta _{1,1}-2\right)
xH_{f}(x)G(x).  \label{dd}
\end{eqnarray}%
where 
\begin{equation}
\Delta (x)=\frac{\beta _{2,1}\beta _{1,N}}{\beta _{1,N-1}}\left( \frac{x-1}{%
1-\epsilon q\sqrt{\zeta }}\right) G(x)
\end{equation}%
The diagonal entries (\ref{dd}) depend on the variables $\beta _{\alpha
,\alpha }$ which are related to the free parameters $\beta _{1,2},\beta
_{2,1},\beta _{1,N-1}$ and $\beta _{1,N}$ through the expressions 
\begin{eqnarray}
\beta _{2,2} &=&\beta _{1,1}-\frac{2\epsilon }{q^{\frac{1}{2}r}-\epsilon }-%
\frac{\beta _{2,1}\beta _{1,N}}{\beta _{1,N-1}}(1-\epsilon q\sqrt{\zeta }), 
\notag \\
\beta _{3,3} &=&\beta _{2,2}+\frac{\beta _{2,1}\beta _{1,N}}{\beta _{1,N-1}},
\notag \\
\beta _{N-1,N-1} &=&\beta _{1,1}-\frac{2\epsilon }{q^{\frac{1}{2}r}-\epsilon 
}+\frac{\beta _{2,1}\beta _{1,N}}{\beta _{1,N-1}}(q^{\frac{1}{2}r}+\epsilon
)q\sqrt{\zeta },  \notag \\
\beta _{N,N} &=&\beta _{1,1}+2+\frac{\beta _{2,1}\beta _{1,N}}{\beta _{1,N-1}%
}(q^{2}+1)(\epsilon q\sqrt{\zeta }).
\end{eqnarray}%
where%
\begin{equation}
\beta _{2,1}=\epsilon \left( \frac{\beta _{1,2}\beta _{1,N-1}}{\beta _{1,N}}%
\frac{1}{q\sqrt{\zeta }}-\frac{2}{(q^{\frac{1}{2}r}-\epsilon )(1-\epsilon q%
\sqrt{\zeta })}\right) \frac{\beta _{1,N-1}}{\beta _{1,N}}
\end{equation}%
Therefore we have a solution with three free parameters.

\subsubsection{\textbf{Solution $\mathcal{M}_{3}$}}

This class of solution is valid for all $U_{q}[osp(r|2m)^{(1)}]$ vertex
models with $m\geq 3$ and the corresponding $K$-matrix possess the following
general form 
\begin{equation}
K^{-}(x)\!=\!\!\left( 
\begin{array}{ccccccc}
k_{1,1}(x) & \cdots & k_{1,m}(x) &  & k_{1,r+m+1}(x) & \cdots & k_{1,N}(x)
\\ 
\vdots & \ddots & \vdots & \mathbb{O}_{m\times r} & \vdots & \ddots & \vdots
\\ 
k_{m,1}(x) & \cdots & k_{m,m}(x) &  & k_{m,r+m+1}(x) & \cdots & k_{m,N}(x)
\\ 
& \!\mathbb{O}_{r\times m} &  & \!\mathbb{K}_{3}(x) &  & \!\mathbb{O}%
_{r\times m} &  \\ 
k_{r+m+1,1}(x) & \cdots & k_{r+m+1,m}(x) &  & k_{r+m+1,r+m+1}(x) & \cdots & 
k_{r+m+1,N}(x) \\ 
\vdots & \ddots & \vdots & \mathbb{O}_{m\times r} & \vdots & \ddots & \vdots
\\ 
k_{N,1}(x) & \cdots & k_{N,m}(x) &  & k_{N,r+m+1}(x) & \cdots & k_{N,N}(x)%
\end{array}%
\right) ,
\end{equation}%
where $\mathbb{K}_{3}(x)$ is a diagonal matrix given by 
\begin{equation}
\mathbb{K}_{3}(x)=k_{m+1,m+1}(x)\;\mathbb{I}_{r\times r}.
\end{equation}

With respect to the elements of the last column, we have the following
expression 
\begin{eqnarray}
k_{i,N}(x) &=&-\frac{\epsilon }{\sqrt{\zeta }}\frac{\theta _{i}q^{t_{i}}}{%
\theta _{1}q^{t_{1}}}\beta _{1,i^{\prime }}xG(x) \\
\;i &=&2,\dots ,m\;\;\mathrm{and\quad }i=r+m+1,\dots ,N-1,  \notag
\end{eqnarray}%
In their turn the entries of the first column are mainly given by 
\begin{eqnarray}
k_{i,1}(x) &=&\frac{\theta _{i}q^{t_{i}}}{\theta _{2}q^{t_{2}}}\frac{\beta
_{2,1}\beta _{1,i^{\prime }}}{\beta _{1,N-1}}G(x), \\
i &=&3,\dots ,m\;\;\mathrm{and\quad }i=r+m+1,\dots ,N-1.  \notag
\end{eqnarray}%
In the last row we have 
\begin{eqnarray}
k_{N,j}(x) &=&-\frac{\epsilon }{\sqrt{\zeta }}\frac{\theta _{N}q^{t_{N}}}{%
\theta _{2}q^{t_{2}}}\frac{\beta _{2,1}\beta _{1,j}}{\beta _{1,N-1}}xG(x) \\
j &=&2,\dots ,m\;\;\mathrm{and\quad }j=r+m+1,\dots ,N-1,  \notag
\end{eqnarray}%
while the elements of the first row are $k_{1,j}(x)=\beta _{1,j}G(x)$ for $%
j=2,\dots ,m$ and\ $j=r+m+1,\dots ,N-1$.

Concerning the elements of the secondary diagonal, they are given by 
\begin{eqnarray}
k_{i,i^{\prime }}(x) &=&-q^{2}\frac{\theta _{1}q^{t_{1}}}{\theta _{i^{\prime
}}q^{t_{i^{\prime }}}}\frac{(1-\epsilon q\sqrt{\zeta })^{2}}{(q+1)^{2}}\frac{%
\beta _{1,i^{\prime }}^{2}}{\beta _{1,N}}G(x)H_{f}(x)  \notag \\
i &=&2,\dots ,m\;,\;i\neq i^{\prime }\quad \mathrm{and\quad }i=r+m+1,\dots
,N-1,
\end{eqnarray}%
while the remaining entries $k_{1,N}(x)$ and $k_{N,1}(x)$ are determined by
the following expressions 
\begin{eqnarray}
k_{1,N}(x) &=&\beta _{1,N}G(x)H_{f}(x)  \notag \\
k_{N,1}(x) &=&\frac{\theta _{N-1}q^{t_{N-1}}}{\theta _{2}q^{t_{2}}}\frac{%
\beta _{1,N}\beta _{2,1}^{2}}{\beta _{1,N-1}^{2}}G(x)H_{f}(x)
\end{eqnarray}

The remaining matrix elements $k_{i,j}(x)$ with $i\neq j$ are then%
\begin{equation}
k_{i,j}(x)=\left\{ 
\begin{array}{c}
-\frac{\epsilon }{\sqrt{\zeta }}\frac{\theta _{i}q^{t_{i}}}{\theta
_{1}q^{t_{1}}}\left( \frac{1-\epsilon q\sqrt{\zeta }}{q+1}\right) \frac{%
\beta _{1,i^{\prime }}\beta _{1,j}}{\beta _{1,N}}G(x),\quad i<j^{\prime
}\quad 2\leq i,j\leq N-1 \\ 
\frac{1}{\zeta }\frac{\theta _{i}q^{t_{i}}}{\theta _{1}q^{t_{1}}}\left( 
\frac{1-\epsilon q\sqrt{\zeta }}{q+1}\right) \frac{\beta _{1,i^{\prime
}}\beta _{1,j}}{\beta _{1,N}}xG(x),\quad i>j^{\prime }\quad 2\leq i,j\leq N-1%
\end{array}%
\right.
\end{equation}%
and%
\begin{equation*}
k_{2,1}(x)=\frac{2(-1)^{m}(\epsilon q\sqrt{\zeta }-1)q^{m-2}}{(1+\epsilon 
\sqrt{\zeta })(q^{\frac{1}{2}r-1}+(-1)^{m}\epsilon )(q^{\frac{1}{2}%
r}-(-1)^{m}\epsilon )}\frac{\beta _{1,N-1}}{\beta _{1,N}}G(x),
\end{equation*}%
\begin{equation}
k_{1,m}(x)=\frac{2\zeta (q+1)^{2}q^{m-1}}{(1-\epsilon q\sqrt{\zeta }%
)(1+\epsilon \sqrt{\zeta })(q^{\frac{1}{2}r-1}+(-1)^{m}\epsilon )(q^{\frac{1%
}{2}r}-(-1)^{m}\epsilon )}\frac{\beta _{1,N}}{\beta _{1,m^{\prime }}}%
G(x),\quad
\end{equation}%
and the parameters $\beta _{1,j}$ are constrained by the relation 
\begin{equation}
\beta _{1,j}=-\beta _{1,j+1}\frac{\beta _{1,N-j}}{\beta _{1,N+1-j}}%
\;\;\;\;\;\;\;\;\;\;\;\;\;\;\;\;\;j=2,\dots ,m-1.
\end{equation}

With regard to the diagonal matrix elements, they are given by 
\begin{equation}
k_{i,i}(x)=\left\{ 
\begin{array}{c}
k_{1,1}(x)+(\beta _{i,i}-\beta _{1,1})G(x),\quad 
\hfill%
2\leq i\leq m \\ 
k_{1,1}(x)+(\beta _{m+1,m+1}-\beta _{1,1})G(x)+\Delta (x),\quad 
\hfill%
i=m+1 \\ 
k_{m+1,m+1}(x)+(\beta _{r+m+1,r+m+1}-\beta _{m+1,m+1})xG(x)+\epsilon q^{r-1}%
\sqrt{\zeta }\Delta (x), \\ 
\hfill%
i=r+m+1 \\ 
k_{i-1,i-1}(x)+(\beta _{i,i}-\beta _{i-1,i-1})xG(x),\quad 
\hfill%
r+m+2\leq i\leq N%
\end{array}%
\right.  \label{rr1}
\end{equation}%
The last term of the recurrence relation (\ref{rr1}) is identified with%
\begin{equation}
k_{N,N}(x)=x^{2}k_{1,1}(x)+(\beta _{N,N}-\beta _{1,1}-2)xG(x)H_{f}(x)
\end{equation}%
to find%
\begin{eqnarray}
k_{1,1}(x) &=&[(\beta _{N,N}-\beta _{m+1,m+1})x-(\beta _{N,N}-\beta
_{1,1}-2)xH_{f}(x)+\beta _{m+1,m+1}-\beta _{1,1}]\frac{G(x)}{x^{2}-1}  \notag
\\
&&+[1+\epsilon q^{r-1}\sqrt{\zeta }]\frac{\Delta (x)}{x^{2}-1}
\end{eqnarray}%
where 
\begin{equation}
\Delta (x)=-\frac{2(x-1)G(x)}{(1+\epsilon \sqrt{\zeta })(q^{\frac{1}{2}%
r-1}+(-1)^{m}\epsilon )(q^{\frac{1}{2}r}-(-1)^{m}\epsilon )}.
\end{equation}%
In their turn the diagonal parameters $\beta _{i,i}$ are fixed by the
relations 
\begin{equation}
\beta _{i,i}=\left\{ 
\begin{array}{c}
\beta _{1,1}+\Lambda _{m}\sum\limits_{k=0}^{i-2}(-\frac{1}{q})^{k},\quad 
\hfill%
2\leq i\leq m \\ 
\beta _{r+m+1,r+m+1}-(-1)^{m}\epsilon q^{\frac{1}{2}r}\Lambda
_{m}\sum\limits_{k=0}^{i-r-m-2}(-\frac{1}{q})^{k},\quad 
\hfill%
r+m+2\leq i\leq N%
\end{array}%
\right.
\end{equation}%
and%
\begin{equation}
\beta _{m+1,m+1}=\beta _{1,1}+\Lambda _{m}\left( \frac{q}{q+1}+(-1)^{m}\frac{%
q^{2-m}}{(q+1)^{2}}\frac{1+\epsilon \sqrt{\zeta }}{\epsilon \sqrt{\zeta }}%
\right) ,
\end{equation}%
\begin{equation}
\beta _{r+m+1,r+m+1}=\beta _{m+1,m+1}+\frac{2q^{r-2}(\epsilon q\sqrt{\zeta }%
-1)}{(1+\epsilon \sqrt{\zeta })(q^{\frac{1}{2}r-1}+(-1)^{m}\epsilon )(q^{%
\frac{1}{2}r}-(-1)^{m}\epsilon )},
\end{equation}%
with 
\begin{equation}
\Lambda _{m}=-\frac{2(-1)^{m}q^{m-2}(1+q)^{2}\epsilon \sqrt{\zeta }}{%
(1+\epsilon \sqrt{\zeta })(q^{\frac{1}{2}r-1}+(-1)^{m}\epsilon )(q^{\frac{1}{%
2}r}-(-1)^{m}\epsilon )}.
\end{equation}%
The class of solution $\mathcal{M}_{3}$ has a total amount of $m$ free
parameters namely $\beta _{1,r+m+1},\dots ,\beta _{1,N}$.

Here we note that we can take the limit $m=2$ in the class $\mathcal{M}_{3}$
to get a two parameter solution which is a particular case of the three
parameter solution of the class $\mathcal{M}_{2}$.

\subsection{Bosonic Solutions}

Here we will turn off the fermionic degree of freedom, i.e. $k_{i,j}(x)=0,$
for $i\neq j=\{1,...,m\}$ and $i\neq j=$ $\left\{ r+m+1,...,N\right\} $ in
order to get $K$-matrix solutions with only bosonic degree of freedom. We
have found five classes of non-diagonal solutions that we refer in what
follows as solutions of type $\mathcal{M}_{4}$ to type $\mathcal{M}_{8}$:

\subsubsection{\textbf{Solution $\mathcal{M}_{4}$:}}

This family of solutions is valid only for the $U_{q}[osp(2|2m)^{(1)}]$
vertex model with $m\geq 1$ and the corresponding $K$-matrix has the
following block diagonal structure%
\begin{equation}
K^{-}(x)=\left( 
\begin{array}{ccc}
k_{1,1}(x)\mathbb{I}_{m\times n} & \mathbb{O}_{m\times 2} & \mathbb{O}%
_{m\times m} \\ 
\mathbb{O}_{2\times m} & 
\begin{array}{cc}
k_{m+1,m+1}(x) & k_{m+1,m+2}(x) \\ 
k_{m+2,m+1}(x) & k_{m+2,m+2}(x)%
\end{array}
& \mathbb{O}_{2\times m} \\ 
\mathbb{O}_{m\times m} & \mathbb{O}_{m\times 2} & k_{N,N}(x)\mathbb{I}%
_{m\times n}%
\end{array}%
\right)
\end{equation}%
The non-null entries are given by 
\begin{eqnarray}
k_{1,1}(x)
&=&1\;,\;\;\;\;\;\;\;\;\;\;\;\;\;\;\;\;\;\;\;\;\;\;\;\;\;\;\;\;\;\;\;%
\;k_{N,N}(x)=x^{2},  \notag \\
k_{m+1,m+1}(x) &=&\frac{x^{2}(1-\zeta )}{x^{2}-\zeta },\;\;\;\;\;\;\;\;\;\;%
\;\;\;\;\;\;\;\;\;\;k_{m+1,m+2}(x)=\frac{\beta }{2}\frac{x(x^{2}-1)(1-\zeta )%
}{x^{2}-\zeta },  \notag \\
k_{m+2,m+1}(x) &=&\frac{2}{\beta }\frac{x(x^{2}-1)\zeta }{(1-\zeta
)(x^{2}-\zeta )},\;\;\;\;\;\;\;\;k_{m+2,m+2}(x)=\frac{x^{2}(1-\zeta )}{%
x^{2}-\zeta }.
\end{eqnarray}%
where $\beta =\beta _{m+1,m+2}$ is a free parameter.

\subsubsection{\textbf{Solution $\mathcal{M}_{5}$:}}

The family $\mathcal{M}_{5}$ is acceptable by the vertex model $%
U_{q}[osp(3|2m)^{(1)}]$ and it is characterized by a $K$-matrix of the form 
\begin{equation}
K^{-}(x)=\left( 
\begin{array}{ccc}
k_{1,1}(x)\mathbb{I}_{m\times m} & \mathbb{O}_{m\times 3} & \mathbb{O}%
_{m\times m} \\ 
\mathbb{O}_{3\times m} & 
\begin{array}{ccc}
k_{m+1,m+1}(x) & k_{m+1,m+2}(x) & k_{m+1,m+3}(x) \\ 
k_{m+2,m+1}(x) & k_{m+2,m+2}(x) & k_{m+2,m+3}(x) \\ 
k_{m+3,m+1}(x) & k_{m+3,m+2}(x) & k_{m+3,m+3}(x)%
\end{array}
& \mathbb{O}_{3\times m} \\ 
\mathbb{O}_{m\times m} & \mathbb{O}_{m\times 3} & k_{N,N}(x)\mathbb{I}%
_{m\times m}%
\end{array}%
\right) .
\end{equation}%
The non-diagonal matrix elements are given by the following expressions 
\begin{eqnarray}
k_{m+1,m+2}(x) &=&\beta
_{m+1,m+2}G(x)\;\;\;\;\;\;\;\;\;\;\;\;\;\;\;k_{m+2,m+1}(x)=\beta
_{m+2,m+1}G(x)  \notag \\
k_{m+3,m+2}(x) &=&\beta
_{m+3,m+2}xG(x)\;\;\;\;\;\;\;\;\;\;\;\;\;k_{m+2,m+3}(x)=\beta _{m+2,m+3}xG(x)
\\
k_{m+1,m+3}(x) &=&\beta
_{m+1,m+3}G(x)H_{b}(x)\;\;\;\;\;\;\;k_{m+3,m+1}(x)=\beta
_{m+3,m+1}G(x)H_{b}(x)  \notag
\end{eqnarray}%
where $H_{b}(x)$ is given by (\ref{Hs}). In their turn the above parameters $%
\beta _{i,j}$ are constrained by the relations 
\begin{eqnarray}
\beta _{m+2,m+3} &=&\epsilon \frac{\sqrt{q}}{\sqrt{\zeta }}\beta
_{m+1,m+2},\quad \beta _{m+3,m+2}=\epsilon \frac{\sqrt{q}}{\sqrt{\zeta }}%
\beta _{m+2,m+1},  \notag \\
\beta _{m+3,m+1} &=&\beta _{m+1,m+3}\left( \frac{\beta _{m+2,m+1}}{\beta
_{m+1,m+2}}\right) ^{2},  \notag \\
\beta _{m+2,m+1} &=&-\epsilon \frac{\sqrt{q}}{\sqrt{\zeta }}\frac{\beta
_{m+1,m+2}}{\beta _{m+1,m+3}}\left[ \frac{\sqrt{q}}{(q+1)}\frac{\beta
_{m+1,m+2}^{2}}{\beta _{m+1,m+3}}-\frac{2\zeta }{(q+\epsilon \sqrt{\zeta }%
)(\epsilon \sqrt{\zeta }-1)}\right] .
\end{eqnarray}

The diagonal entries are then given by 
\begin{eqnarray}
k_{1,1}(x) &=&\left[ \frac{2}{x^{2}-1}-\frac{\epsilon (qx+\zeta )}{\sqrt{%
q\zeta }x(x+1)}\frac{\beta _{m+1,m+3}\beta _{m+2,m+1}}{\beta _{m+1,m+2}}%
\right] G(x)H_{b}(x)  \notag \\
&&-\frac{\epsilon \sqrt{q}}{\sqrt{\zeta }}\frac{\beta _{m+1,m+2}^{2}}{\beta
_{m+1,m+3}}\frac{G(x)}{x+1}  \notag \\
k_{m+1,m+1}(x) &=&\left[ \frac{2}{x^{2}-1}-\frac{\epsilon (q-\zeta )}{\sqrt{%
q\zeta }(x+1)}\frac{\beta _{m+1,m+3}\beta _{m+2,m+1}}{\beta _{m+1,m+2}}%
\right] G(x)H_{b}(x)  \notag \\
&&-\frac{\epsilon \sqrt{q}}{\sqrt{\zeta }}\frac{\beta _{m+1,m+2}^{2}}{\beta
_{m+1,m+3}}\frac{G(x)}{x+1}  \notag \\
k_{m+2,m+2}(x) &=&k_{m+1,m+1}(x)+\left( \beta _{m+2,m+2}-\beta
_{m+1,m+1}\right) G(x)+\Delta (x)  \notag \\
k_{m+3,m+3}(x) &=&k_{m+2,m+2}(x)+\left( \beta _{m+3,m+3}-\beta
_{m+2,m+2}\right) xG(x)+\epsilon \sqrt{\zeta }\Delta (x)  \notag \\
k_{N,N}(x) &=&x^{2}k_{1,1}(x)
\end{eqnarray}%
where 
\begin{equation}
\Delta (x)=-\frac{\sqrt{q}}{(q+\epsilon \sqrt{\zeta })}\frac{\beta
_{m+2,m+1}\beta _{m+1,m+3}}{\beta _{m+1,m+2}}(x-1)G(x),
\end{equation}%
and the parameters $\beta _{m+1,m+1},\beta _{m+2,m+2}$ and $\beta _{m+3,m+3}$
are fixed by the relations 
\begin{eqnarray}
\beta _{m+1,m+1} &=&\beta _{1,1}+\frac{\epsilon \sqrt{\zeta }}{\sqrt{q}}%
\frac{\beta _{m+1,m+3}\beta _{m+2,m+1}}{\beta _{m+1,m+2}}  \notag \\
\beta _{m+2,m+2} &=&\beta _{1,1}+\frac{2\epsilon \sqrt{\zeta }}{(q+\epsilon 
\sqrt{\zeta })}+\epsilon \frac{\sqrt{\zeta }}{\sqrt{q}}\frac{\beta
_{m+1,m+2}^{2}}{\beta _{m+1,m+3}} \\
\beta _{m+3,m+3} &=&\beta _{1,1}+2+\frac{\epsilon \sqrt{q}}{\sqrt{\zeta }}%
\frac{\beta _{m+1,m+3}\beta _{m+2,m+1}}{\beta _{m+1,m+2}}.  \notag
\end{eqnarray}%
The solution $\mathcal{M}_{5}$ possess two free parameters namely $\beta
_{m+1,m+2}$ and $\beta _{m+1,m+3}$.

\subsubsection{\textbf{Solution $\mathcal{M}_{6}$:}}

The solution $\mathcal{M}_{6}$ is admitted for the $U_{q}[osp(4|2m)^{(1)}]$
models with the following $K$-matrix 
\begin{eqnarray}
&&K^{-}(x)=\!  \notag \\
&&\left( 
\begin{array}{ccc}
k_{1,1}(x)\mathbb{I}_{m\times m}\! & \mathbb{O}_{m\times 4}\! & \mathbb{O}%
_{m\times m}\! \\ 
\mathbb{O}_{4\times m}\! & 
\begin{array}{cccc}
k_{m+1,m+1}(x) & k_{m+1,m+2}(x) & k_{m+1,m+3}(x) & k_{m+1,m+4}(x) \\ 
k_{m+2,m+1}(x) & k_{m+2,m+2}(x) & k_{m+2,m+3}(x) & k_{m+2,m+4}(x) \\ 
k_{m+3,m+1}(x) & k_{m+3,m+2}(x) & k_{m+3,m+3}(x) & k_{m+3,m+4}(x) \\ 
k_{m+4,m+1}(x) & k_{m+4,m+2}(x) & k_{m+4,m+3}(x) & k_{m+4,m+4}(x)%
\end{array}%
\! & \mathbb{O}_{4\times m}\! \\ 
\mathbb{O}_{m\times m\!}\! & \mathbb{O}_{m\times 4}\! & k_{N,N}(x)\mathbb{I}%
_{m\times m}\!%
\end{array}%
\right) .  \notag \\
&&
\end{eqnarray}%
The non-diagonal elements are all grouped in the $4\times 4$ central block
matrix. With respect to this central block, the entries of the secondary
diagonal are given by 
\begin{eqnarray}
k_{m+2,m+3}(x) &=&-\frac{\beta _{m+2,m+1}}{\beta _{m+1,m+2}}k_{m+1,m+4}(x) 
\notag \\
k_{m+3,m+2}(x) &=&-\frac{q^{2}}{\zeta }\frac{\beta _{m+1,m+2}\Gamma _{m}^{2}%
}{\beta _{m+2,m+1}\beta _{m+1,m+4}^{2}}k_{m+1,m+4}(x) \\
k_{m+4,m+1}(x) &=&\frac{q^{2}}{\zeta }\frac{\Gamma _{m}^{2}}{\beta
_{m+1,m+4}^{2}}k_{m+1,m+4}(x),  \notag
\end{eqnarray}%
and the remaining non-diagonal elements can be written as 
\begin{eqnarray}
k_{m+1,m+2}(x) &=&\beta
_{m+1,m+2}G_{1}(x)\;\;\;\;\;\;\;\;\;\;\;\;\;\;\;\;\;\;\;\;\;\;\;\;%
\;k_{m+2,m+1}(x)=\beta _{m+2,m+1}G_{1}(x)  \notag \\
k_{m+1,m+3}(x) &=&\beta
_{m+1,m+3}G_{2}(x)\;\;\;\;\;\;\;\;\;\;\;\;\;\;\;\;\;\;\;\;\;\;\;\;%
\;k_{m+3,m+1}(x)=\frac{q^{2}}{\zeta }\frac{\beta _{m+1,m+2}\beta
_{m+1,m+3}\Gamma _{m}^{2}}{\beta _{m+2,m+1}\beta _{m+1,m+4}^{2}}G_{2}(x) 
\notag \\
k_{m+2,m+4}(x) &=&-\frac{\beta _{m+2,m+1}\beta _{m+1,m+4}}{\Gamma _{m}}%
xG_{1}(x)\;\;\;\;\;\;\;\;\;\;\;\;\;\;\;k_{m+4,m+2}(x)=-\frac{q^{2}}{\zeta }%
\frac{\beta _{m+1,m+2}\Gamma _{m}}{\beta _{m+1,m+4}}xG_{1}(x)  \notag \\
k_{m+3,m+4}(x) &=&-\frac{q^{2}}{\zeta }\frac{\beta _{m+1,m+2}\beta
_{m+1,m+3}\Gamma _{m}}{\beta _{m+1,m+4}\beta _{m+2,m+1}}xG_{2}(x)\;\;\;\;\;%
\;\;k_{m+4,m+3}(x)=-\frac{q^{2}}{\zeta }\frac{\beta _{m+1,m+3}\Gamma _{m}}{%
\beta _{m+1,m+4}}xG_{2}(x),  \notag \\
&&
\end{eqnarray}%
where 
\begin{eqnarray}
G_{1}(x) &=&\left[ \frac{\zeta -q^{2}x}{q^{2}(x-1)}+\frac{\beta
_{m+1,m+3}\Gamma _{m}}{\beta _{m+2,m+1}\beta _{m+1,m+4}}\right] \frac{%
q^{2}(x-1)}{(\zeta -q^{2}x^{2})}\frac{k_{m+1,m+4}(x)}{\beta _{m+1,m+4}}, 
\notag \\
G_{2}(x) &=&\left[ \frac{\zeta -q^{2}x}{x-1}+\frac{\zeta \beta
_{m+2,m+1}\beta _{m+1,m+4}}{\beta _{m+1,m+3}\Gamma _{m}}\right] \frac{(x-1)}{%
(\zeta -q^{2}x^{2})}\frac{k_{m+1,m+4}(x)}{\beta _{m+1,m+4}}
\end{eqnarray}%
and 
\begin{equation}
\Gamma _{m}=\frac{\beta _{m+1,m+2}\beta _{m+1,m+3}}{\beta _{m+1,m+4}}+\frac{%
2\zeta }{q^{2}-\zeta }.
\end{equation}%
In their turn the diagonal entries are given by the following expressions 
\begin{eqnarray}
k_{1,1}(x) &=&\left\{ \frac{\zeta -q^{2}x}{(x+1)(\zeta -q^{2}x^{2})}\left[ 
\frac{\beta _{m+1,m+2}\beta _{m+2,m+1}}{\Gamma _{m}}\right. \right.  \notag
\\
&&+\left. \left. \frac{\beta _{m+1,m+2}\beta _{m+1,m+3}}{\zeta \beta
_{m+1,m+4}}\left( \frac{q^{2}\beta _{m+1,m+3}\Gamma _{m}}{\beta
_{m+1,m+4}\beta _{m+2,m+1}}+\frac{(\zeta +q^{2}x^{2})}{x}\right) \right]
\right.  \notag \\
&&+\left. \frac{2\left( \zeta -q^{2}x^{2}\right) }{x\left( x^{2}-1\right)
\left( \zeta -q^{2}\right) }\right\} \frac{k_{m+1,m+4}(x)}{\beta _{m+1,m+4}}
\end{eqnarray}%
with the recurrence relation%
\begin{eqnarray}
k_{m+1,m+1}(x) &=&k_{1,1}(x)+(\beta _{m+1,m+1}-\beta _{1,1})G_{1}(x)+\Delta
_{1}(x),  \notag \\
k_{m+2,m+2}(x) &=&k_{m+1,m+1}(x)+(\beta _{m+2,m+2}-\beta _{m+1,m+1})G_{1}(x),
\notag \\
k_{m+3,m+3}(x) &=&k_{m+2,m+2}(x)+\Delta _{2}(x),  \notag \\
k_{m+4,m+4}(x) &=&k_{m+3,m+3}(x)+(\beta _{m+4,m+4}-\beta
_{m+3,m+3})xG_{2}(x),  \notag \\
k_{N,N}(x) &=&x^{2}k_{1,1}(x),
\end{eqnarray}%
where 
\begin{eqnarray}
\Delta _{1}(x) &=&\frac{\Gamma _{m}\left( \Gamma _{m}\beta
_{m+1,m+3}q^{2}x+\zeta \beta _{m+1,m+4}\beta _{m+2,m+1}\right) }{\beta
_{m+1,m+4}^{2}\beta _{m+2,m+1}}\frac{(x-1)k_{m+1,m+4}(x)}{x(\zeta
-q^{2}x^{2})}  \notag \\
\Delta _{2}(x) &=&\frac{\beta _{m+1,m+2}\left( \zeta \beta
_{m+2,m+1}^{2}\beta _{m+1,m+4}^{2}-q^{2}\beta _{m+1,m+3}^{2}\Gamma
_{m}^{2}\right) }{\Gamma _{m}\beta _{m+1,m+4}^{3}\beta _{m+2,m+1}}\frac{%
x(\zeta -q^{2})k_{m+1,m+4}(x)}{\zeta (\zeta -q^{2}x^{2})}.  \notag \\
&&
\end{eqnarray}%
The variables $\beta _{i,j}$ are given in terms of the free parameters $%
\beta _{m+1,m+2},\beta _{m+2,m+1},\beta _{m+1,m+3}$ and $\beta _{m+1,m+4}$
through the relations \ here 
\begin{eqnarray}
\beta _{m+2,m+2} &=&\beta _{1,1}+\frac{2\zeta }{\zeta -q^{2}}-\frac{\beta
_{m+1,m+2}\beta _{m+2,m+1}}{\Gamma _{m}}  \notag \\
\beta _{m+3,m+3} &=&\beta _{1,1}+\frac{2\zeta }{\zeta -q^{2}}-\frac{q^{2}}{%
\zeta }\frac{\Gamma _{m}\beta _{m+1,m+2}\beta _{m+1,m+3}^{2}}{\beta
_{m+2,m+1}\beta _{m+1,m+4}^{2}}  \notag \\
\beta _{m+4,m+4} &=&\beta _{1,1}+\frac{2\zeta }{\zeta -q^{2}}-\frac{q^{2}}{%
\zeta }\frac{\beta _{m+1,m+2}\beta _{m+1,m+3}}{\beta _{m+1,m+4}}.
\end{eqnarray}%
We remark here that the form of this class of solution differs from general
form (\ref{Gf}), which is recovered by the condition $G_{1}(x)=G_{2}(x)$,
but reducing the number of free parameters.

\subsubsection{\textbf{Solution $\mathcal{M}_{7}$:}}

The vertex model $U_{q}[osp(2n|2m)^{(1)}]$ admits the solution $\mathcal{M}%
_{7}$ for $n\geq 3$, whose $K$-matrix has the following block structure 
\begin{equation}
K^{-}(x)=\left( 
\begin{array}{ccc}
k_{1,1}(x)\mathbb{I}_{m\times m} & \mathbb{O}_{m\times 2n} & \mathbb{O}%
_{m\times m} \\ 
\mathbb{O}_{2n\times m} & 
\begin{array}{ccc}
k_{m+1,m+1}(x) & \cdots & k_{m+1,2n+m}(x) \\ 
\vdots & \ddots & \vdots \\ 
k_{2n+m,m+1}(x) & \cdots & k_{2n+m,2n+m}(x)%
\end{array}
& \mathbb{O}_{2n\times m} \\ 
\mathbb{O}_{m\times m} & \mathbb{O}_{m\times 2n} & k_{N,N}(x)\mathbb{I}%
_{m\times m}%
\end{array}%
\right)
\end{equation}

The central block matrix cluster all non-diagonal elements different from
zero. Concerning that central block, we have the following expressions
determining entries of the borders, 
\begin{eqnarray}
k_{i,2n+m}(x) &=&\frac{\epsilon }{\sqrt{\zeta }}q^{t_{i}-t_{m+1}}\beta
_{m+1,i^{\prime
}}xG(x),\;\;\;\;\;\;\;\;\;\;\;\;\;\;\;\;\;\;\;\;\;\;\;\;\;\;\;\;i=m+2,\dots
,2n+m-1  \notag \\
k_{2n+m,j}(x) &=&\frac{\epsilon }{\sqrt{\zeta }}q^{t_{2n+m}-t_{m+2}}\frac{%
\beta _{m+2,m+1}\beta _{m+1,j}}{\beta _{m+1,2n+m-1}}xG(x),\;\;\;\;\;\;%
\;j=m+2,\dots ,2n+m-1  \notag \\
k_{i,m+1}(x) &=&q^{t_{i}-t_{m+2}}\frac{\beta _{m+2,m+1}\beta _{m+1,i^{\prime
}}}{\beta _{m+1,2n+m-1}}G(x),\;\;\;\;\;\;\;\;\;\;\;\;\;\;\;\;\;\;\;\;i=m+3,%
\dots ,2n+m-1  \notag \\
k_{m+1,j}(x) &=&\beta
_{m+1,j}G(x).\;\;\;\;\;\;\;\;\;\;\;\;\;\;\;\;\;\;\;\;\;\;\;\;\;\;\;\;\;\;\;%
\;\;\;\;\;\;\;\;\;\;\;\;\;\;\;\;\;\;j=m+2,\dots ,2n+m-1
\end{eqnarray}%
The entries of the secondary diagonal are given by 
\begin{equation}
k_{i,i^{\prime }}(x)=\left\{ 
\begin{array}{c}
\displaystyle%
\beta _{m+1,2n+m}G(x)H_{b}(x),\qquad \qquad \qquad \qquad \qquad \qquad
\qquad \qquad \qquad \qquad i=m+1 \\ 
\displaystyle%
q^{2m}q^{t_{m+1}-t_{i^{\prime }}}\left( \frac{q+\epsilon \sqrt{\zeta }}{q+1}%
\right) ^{2}\frac{\beta _{m+1,i^{\prime }}^{2}}{\beta _{m+1,2n+m}}%
G(x)H_{b}(x),\qquad i=m+2,...,2n+m-1 \\ 
\displaystyle%
q^{t_{2n+m-1}-t_{m+2}}\frac{\beta _{m+2,m+1}^{2}\beta _{m+1,2n+m}}{\beta
_{m+1,2n+m-1}^{2}}G(x)H_{b}(x),\qquad \qquad \qquad \qquad \qquad i=2n+m%
\end{array}%
\right.
\end{equation}%
and the remaining non-diagonal elements are determined by the expression 
\begin{equation}
k_{i,j}(x)=\left\{ 
\begin{array}{c}
\displaystyle%
\frac{\epsilon }{\sqrt{\zeta }}q^{t_{i}-t_{m+1}}\left( \frac{q+\epsilon 
\sqrt{\zeta }}{q+1}\right) \frac{\beta _{m+1,i^{\prime }}\beta _{m+1,j}}{%
\beta _{m+1,2n+m}}G(x),\qquad i<j^{\prime },\ m+1<i,j<2n+m \\ 
\\ 
\displaystyle%
\frac{1}{\zeta }q^{t_{i}-t_{m+1}}\left( \frac{q+\epsilon \sqrt{\zeta }}{q+1}%
\right) \frac{\beta _{m+1,i^{\prime }}\beta _{m+1,j}}{\beta _{m+1,2n+m}}%
xG(x),\qquad i>j^{\prime },\ m+1<i,j<2n+m%
\end{array}%
\right. .
\end{equation}%
and%
\begin{eqnarray}
k_{m+1,m+n}(x) &=&\frac{2\epsilon \sqrt{\zeta }}{(1-\epsilon \sqrt{\zeta }%
)(q+\epsilon \sqrt{\zeta })}\frac{(-1)^{n}(1+q)^{2}\zeta }{%
(q^{n}-(-1)^{n}\epsilon \sqrt{\zeta })(q^{n-1}-(-1)^{n}\epsilon \sqrt{\zeta }%
)}\frac{\beta _{m+1,2n+m}}{\beta _{m+1,n+m+1}}G(x),  \notag \\
k_{m+2,m+1}(x) &=&-\frac{2\epsilon \sqrt{\zeta }}{(1-\epsilon \sqrt{\zeta })}%
\frac{q(q+\epsilon \sqrt{\zeta })}{(q^{n}-(-1)^{n}\epsilon \sqrt{\zeta }%
)(q^{n-1}-(-1)^{n}\epsilon \sqrt{\zeta })}\frac{\beta _{m+1,2n+m-1}}{\beta
_{m+1,2n+m}}G(x).
\end{eqnarray}

In their turn the diagonal entries $k_{i,i}(x)$ are given by%
\begin{equation}
k_{1,1}(x)=\left( \frac{x-\epsilon \sqrt{\zeta }}{1-\epsilon \sqrt{\zeta }}%
\right) \left( \frac{xq^{n}-(-1)^{n}\epsilon \sqrt{\zeta }}{%
q^{n}-(-1)^{n}\epsilon \sqrt{\zeta }}\right) \left( \frac{%
xq^{n-1}-(-1)^{n}\epsilon \sqrt{\zeta }}{q^{n-1}-(-1)^{n}\epsilon \sqrt{%
\zeta }}\right) \frac{2G(x)}{x(x^{2}-1)}
\end{equation}%
\begin{equation}
k_{i,i}(x)=\left\{ 
\begin{array}{c}
k_{1,1}(x)+\Gamma _{n}(x),\qquad \qquad \qquad \qquad \qquad 
\hfill%
i=m+1 \\ 
k_{m+1,m+1}(x)+(\beta _{i,i}-\beta _{m+1,m+1})G(x),\qquad 
\hfill%
i=m+2,...,m+n \\ 
k_{n+m,n+m}(x),\qquad \qquad \qquad \qquad \qquad \qquad 
\hfill%
i=n+m+1 \\ 
k_{n+m,n+m}(x)+(\beta _{i,i}-\beta _{n+m,n+m})xG(x),\qquad 
\hfill%
i=n+m+2,...,2n+m \\ 
x^{2}k_{1,1}(x),\qquad \qquad \qquad \qquad \qquad \qquad \qquad 
\hfill%
i=N%
\end{array}%
\right.
\end{equation}%
where 
\begin{equation}
\Gamma _{n}(x)=-\frac{2\zeta (qx+\epsilon \sqrt{\zeta })}{(1-\epsilon \sqrt{%
\zeta })(q^{n}-(-1)^{n}\epsilon \sqrt{\zeta })(q^{n-1}-(-1)^{n}\epsilon 
\sqrt{\zeta })}\frac{G(x)}{x}.
\end{equation}

The diagonal parameters $\beta _{i,i}$ are then fixed by the relations 
\begin{equation}
\beta _{i,i}=\left\{ 
\begin{array}{c}
\displaystyle%
\beta _{1,1}-\frac{q+\epsilon \sqrt{\zeta }}{(1+q)^{2}}\Delta _{n},\qquad 
\hfill%
i=m+1 \\ 
\displaystyle%
\beta _{m+1,m+1}+\Delta _{n}\sum_{k=0}^{i-m-2}(-q)^{k},\qquad 
\hfill%
i=m+2,...,n+m \\ 
\beta _{n+m,n+m},\qquad 
\hfill%
i=n+m+1 \\ 
\displaystyle%
\beta _{n+m,n+m}-\epsilon (-1)^{n}q^{2}\Delta
_{n}\sum_{k=0}^{i-n-m-2}(-q)^{k},\qquad 
\hfill%
i=m+2,...,n+m%
\end{array}%
\right.
\end{equation}%
and the auxiliary parameter $\Delta _{n}$ is given by 
\begin{equation}
\Delta _{n}=\frac{2\zeta (1+q)^{2}}{(1-\epsilon \sqrt{\zeta }%
)(q^{n}-(-1)^{n}\epsilon \sqrt{\zeta })(q^{n-1}-(-1)^{n}\epsilon \sqrt{\zeta 
})}.
\end{equation}%
Besides the above relations the following constraints should also holds 
\begin{equation}
\beta _{m+1,m+j}=-\beta _{m+1,j+m+1}\frac{\beta _{m+1,2n+m-j}}{\beta
_{m+1,2n+m+1-j}}\;\;\;\;\;\;\;\;\;\;\;\;\;\;\;\;\;\;\;\;\;\;j=2,\dots ,n-1,
\end{equation}%
and $\beta _{m+1,m+n+1},\dots ,\beta _{m+1,2n+m}$ are regarded as the $n$
free parameters. The case $n=2$ is a particular solution of the three
parameter class $\mathcal{M}_{6}$.

\subsubsection{\textbf{Solution $\mathcal{M}_{8}$:}}

For $n\geq 2$ the vertex model $U_{q}[osp(2n+1|2m)^{(1)}]$ admits the family
of solutions $\mathcal{M}_{8}$ whose $K$-matrix is of the form%
\begin{equation}
K^{-}(x)=\left( 
\begin{array}{ccc}
k_{1,1}(x)\mathbb{I}_{m\times m} & \mathbb{O}_{m\times 2n+1} & \mathbb{O}%
_{m\times m} \\ 
\mathbb{O}_{2n+1\times m} & 
\begin{array}{ccc}
k_{m+1,m+1}(x) & \cdots & k_{m+1,2n+1+m}(x) \\ 
\vdots & \ddots & \vdots \\ 
k_{2n+1+m,m+1}(x) & \cdots & k_{2n+m,2n+m}(x)%
\end{array}
& \mathbb{O}_{2n+1\times m} \\ 
\mathbb{O}_{m\times m} & \mathbb{O}_{m\times 2n+1} & k_{N,N}(x)\mathbb{I}%
_{m\times m}%
\end{array}%
\right)
\end{equation}%
In the central block matrix we find all non-diagonal elements different from
zero similarly to the structure of the solution $\mathcal{M}_{7}$. The
borders of the central block are then determined by the following
expressions 
\begin{eqnarray}
k_{i,2n+m+1}(x) &=&\frac{\epsilon }{\sqrt{\zeta }}q^{t_{i}-t_{m+1}}\beta
_{m+1,i^{\prime
}}xG(x)\;\;\;\;\;\;\;\;\;\;\;\;\;\;\;\;\;\;\;\;\;\;\;\;\;\;\;\;\;\;%
\hfill%
\;i=m+2,\dots ,2n+m  \notag \\
k_{2n+m+1,j}(x) &=&\frac{\epsilon }{\sqrt{\zeta }}q^{t_{2n+m+1}-t_{m+2}}%
\frac{\beta _{m+2,m+1}\beta _{m+1,j}}{\beta _{m+1,2n+m}}xG(x)\;\;\;\;\;\;\;\;%
\hfill%
j=m+2,\dots ,2n+m  \notag \\
k_{i,m+1}(x) &=&q^{t_{i}-t_{m+2}}\frac{\beta _{m+2,m+1}\beta _{m+1,i^{\prime
}}}{\beta _{m+1,2n+m}}G(x)\;\;\;\;\;\;\;\;\;\;\;\;\;\;\;\;\;\;\;\;\;\;%
\hfill%
\;i=m+3,\dots ,2n+m  \notag \\
k_{m+1,j}(x) &=&\beta
_{m+1,j}G(x)\;\;\;\;\;\;\;\;\;\;\;\;\;\;\;\;\;\;\;\;\;\;\;\;\;\;\;\;\;\;\;\;%
\;\;\;\;\;\;\;\;\;\;\;\;\;\;\;\;\;\;\;\;\;%
\hfill%
\;j=m+2,\dots ,2n+m
\end{eqnarray}%
The secondary diagonal elements are given by 
\begin{equation}
k_{i,i^{\prime }}(x)=\left\{ 
\begin{array}{c}
\beta _{m+1,2n+m+1}G(x)H_{b}(x),\qquad 
\hfill%
i=m+1 \\ 
\displaystyle%
q^{2m}q^{t_{m+1}-t_{i^{\prime }}}\left( \frac{q+\epsilon \sqrt{\zeta }}{q+1}%
\right) ^{2}\frac{\beta _{m+1,i^{\prime }}^{2}}{\beta _{m+1,2n+m+1}}%
G(x)H_{b}(x),\qquad 
\hfill%
i=m+2,...,2n+m \\ 
q^{t_{2n+m}-t_{m+2}}\frac{\beta _{m+2,m+1}^{%
{\acute{}}%
2}\beta _{m+1,2n+m+1}}{\beta _{m+1,2n+m}^{2}}G(x)H_{b}(x),\qquad 
\hfill%
i=2n+m+1%
\end{array}%
\right.
\end{equation}

The remaining non-diagonal entries are determined by 
\begin{equation}
k_{i,j}(x)=\left\{ 
\begin{array}{c}
\displaystyle%
\frac{\epsilon }{\sqrt{\zeta }}q^{t_{i}-t_{m+1}}\left( \frac{q+\epsilon 
\sqrt{\zeta }}{q+1}\right) \frac{\beta _{m+1,i^{\prime }}\beta _{m+1,j}}{%
\beta _{m+1,2n+m+1}}G(x),\qquad i<j^{\prime },\ m+2<i,j<2n+m \\ 
\\ 
\displaystyle%
\frac{1}{\zeta }q^{t_{i}-t_{m+1}}\left( \frac{q+\epsilon \sqrt{\zeta }}{q+1}%
\right) \frac{\beta _{m+1,i^{\prime }}\beta _{m+1,j}}{\beta _{m+1,2n+m+1}}%
xG(x),\qquad i>j^{\prime },\ m+2<i,j<2n+m%
\end{array}%
\right. .
\end{equation}%
and%
\begin{eqnarray}
k_{m+2,m+1}(x) &=&-\frac{(-1)^{n}q}{\zeta }\left( \frac{q+\epsilon \sqrt{%
\zeta }}{q+1}\right) ^{2}\frac{\beta _{m+1,n+m}\beta _{m+1,n+m+2}\beta
_{m+1,2n+m}}{\beta _{m+1,2n+m+1}^{2}}G(x),  \notag \\
k_{m+1,m+n}(x) &=&\frac{\epsilon (q+1)(-1)^{n}}{(q^{m+\frac{1}{2}%
}-(-1)^{n}\epsilon )^{2}}\left[ q^{m}\frac{\beta _{m+1,n+m+1}^{2}}{\beta
_{m+1,n+m+2}}+\frac{2(q+1)\sqrt{\zeta }}{(1-\epsilon \sqrt{\zeta }%
)(q+\epsilon \sqrt{\zeta })}\frac{\beta _{m+1,2n+m+1}}{\beta _{m+1,n+m+2}}%
\right] G(x),  \notag \\
&&
\end{eqnarray}%
while the diagonal matrix elements are given by the relations%
\begin{eqnarray}
k_{1,1}(x) &=&\left( \frac{x-\epsilon \sqrt{\zeta }}{1-\epsilon \sqrt{\zeta }%
}\right) \left( \frac{xq^{m+\frac{1}{2}}-(-1)^{n}\epsilon }{q^{m+\frac{1}{2}%
}-(-1)^{n}\epsilon }\right) ^{2}\frac{2G(x)}{x(x^{2}-1)}  \notag \\
&&+\frac{q^{m}(xq^{2m+1}-1)}{(q+1)\sqrt{\zeta }}\frac{(x-\epsilon \sqrt{%
\zeta })(q+\epsilon \sqrt{\zeta })}{(q^{m+\frac{1}{2}}-(-1)^{n}\epsilon )^{2}%
}\frac{\beta _{m+1,m+n+1}^{2}}{\beta _{m+1,2n+m+1}}\frac{G(x)}{x(x+1)}
\end{eqnarray}%
\begin{equation}
k_{i,i}(x)=\left\{ 
\begin{array}{c}
k_{1,1}(x)+\Gamma (x),\qquad 
\hfill%
i=m+1 \\ 
k_{m+1,m+1}(x)+(\beta _{i,i}-\beta _{m+1,m+1})G(x),\qquad 
\hfill%
i=m+2,...,n+m \\ 
k_{m+1,m+1}(x)+(\beta _{n+m+1,n+m+1}-\beta _{m+1,m+1})G(x)+\Delta (x),\qquad 
\hfill%
i=n+m+1 \\ 
k_{n+m+1,n+m+1}(x)+(\beta _{n+m+2,n+m+2}-\beta _{n+m+1,n+m+1})xG(x)+\epsilon 
\sqrt{\zeta }\Delta (x)%
\hfill%
, \\ 
\hfill%
i=n+m+2 \\ 
k_{i-1,i-1}(x)+(\beta _{i,i}-\beta _{i-1,i-1})xG(x),\qquad 
\hfill%
i=n+m+3,...,2n+m+1 \\ 
x^{2}k_{1,1}(x),\qquad 
\hfill%
i=N%
\end{array}%
\right.
\end{equation}%
The auxiliary functions $\Delta (x)$ and $\Gamma (x)$ are 
\begin{eqnarray}
\Delta (x) &=&-\frac{q^{n}(q+\epsilon \sqrt{\zeta })}{\zeta (q+1)^{2}}\frac{%
\beta _{m+1,n+m}\beta _{m+1,n+m+2}}{\beta _{m+1,2n+m+1}}(x-1)G(x)  \notag \\
\Gamma (x) &=&(-1)^{n+1}\frac{(xq+\epsilon \sqrt{\zeta })(q+\epsilon \sqrt{%
\zeta })}{\epsilon \sqrt{\zeta }x(q+1)^{2}}\frac{\beta _{m+1,n+m}\beta
_{m+1,n+m+2}}{\beta _{m+1,2n+m+1}}G(x),
\end{eqnarray}%
and the parameters $\beta _{m+1,m+j}$ are constrained by the recurrence
formula 
\begin{equation}
\beta _{m+1,j+m}=-\frac{\beta _{m+1,j+m+1}\beta _{m+1,2n+m+1-j}}{\beta
_{m+1,2n+m+2-j}}\;\;\;\;\;\;\;\;\;\;\;\;\;\;\;j=2,\dots ,n-1.
\end{equation}%
In their turn the diagonal parameters $\beta _{i,i}$ are fixed by 
\begin{equation}
\beta _{i,i}=\left\{ 
\begin{array}{c}
\displaystyle%
\beta _{m+1,m+1}+Q_{n,m}\sum_{k=0}^{i-2-m}(-q)^{k},\qquad 
\hfill%
i=m+2,...,n+m \\ 
\displaystyle%
\beta _{n+m+2,n+m+2}-(-1)^{n}q^{2m-n+1}\epsilon \sqrt{\zeta }%
Q_{n,m}\sum_{k=0}^{i-3-n-m}(-q)^{k}, \\ 
\hfill%
i=n+m+3,...,2n+m+1%
\end{array}%
\right.
\end{equation}%
and%
\begin{eqnarray}
\beta _{m+1,m+1} &=&\beta _{1,1}-\epsilon (-1)^{n}q^{m-n+\frac{1}{2}}\left( 
\frac{q+\epsilon \sqrt{\zeta }}{q+1}\right) ^{2}\frac{\beta _{m+1,n+m}\beta
_{m+1,n+m+2}}{\beta _{m+1,2n+m+1}}  \notag \\
\beta _{n+m+1,n+m+1} &=&\beta _{1,1}-\frac{2\epsilon (-1)^{n}}{q^{m+\frac{1}{%
2}}-(-1)^{n}\epsilon }+\frac{q^{m}(1+\epsilon q^{m-n+\frac{3}{2}%
})(q^{n}-(-1)^{n})}{(q+1)(q^{m+\frac{1}{2}}-(-1)^{n}\epsilon )}\frac{\beta
_{m+1,n+m+1}^{2}}{\beta _{m+1,2n+m+1}}  \notag \\
\beta _{n+m+2,n+m+2} &=&\beta _{n+m+1,n+m+1}+\frac{q^{2m+1-n}(q+\epsilon 
\sqrt{\zeta })^{2}}{(q+1)^{2}}\frac{\beta _{m+1,n+m}\beta _{m+1,n+m+2}}{%
\beta _{m+1,2n+m+1}}  \notag \\
&&-\frac{q^{2m-n+\frac{1}{2}}(q+\epsilon \sqrt{\zeta })}{(q+1)}\frac{\beta
_{m+1,n+m+1}^{2}}{\beta _{m+1,2n+m+1}}
\end{eqnarray}%
where 
\begin{equation}
Q_{n,m}=-\frac{2(q+1)^{2}}{(\epsilon \sqrt{\zeta }-1)(q^{m+\frac{1}{2}%
}-(-1)^{n}\epsilon )^{2}}+\frac{q^{m}(q+1)(q+\epsilon \sqrt{\zeta })}{\sqrt{%
\zeta }(q^{m+\frac{1}{2}}-(-1)^{n}\epsilon )^{2}}\frac{\beta _{m+1,n+m+1}^{2}%
}{\beta _{m+1,2n+m+1}}.
\end{equation}

This solution has altogether $n+1$ free parameters corresponding to the set
of variables $\beta _{m+1,n+m+1},\dots ,\beta _{m+1,2n+m+1}$.

\subsection{Complete K-matrices}

The complete $K$-matrices are solutions with all entries different from
zero. This kind of solution will be present only in four class: \ the models
with one or two bosonic degree of freedom, $U_{q}[osp(1|2m)^{(1)}]$ and $%
U_{q}[osp(2|2m)^{(1)}]$ \ respectively and those models with only two
fermionic degree of freedom, $U_{q}[osp(2n|2)^{(1)}]$ and $%
U_{q}[osp(2n+1|2)^{(1)}].$ The special cases $U_{q}[osp(1|2)^{(1)}]$ and $%
U_{q}[osp(2|2)^{(1)}]$ will be presented in appendices.

\subsubsection{\textbf{Solution $\mathcal{M}_{9}$:}}

The family $\mathcal{M}_{9}$ consist of a solution of the reflection
equation where all entries of the $K$-matrix are non-null. This solution is
admitted only by the $U_{q}[osp(1|2m)^{(1)}]$ vertex model. The associated $%
K $-matrix is of the general form (\ref{KM}) and the matrix elements of the
borders are mainly given by 
\begin{eqnarray}
k_{i,N}(x) &=&-\frac{\epsilon }{\sqrt{\zeta }}\frac{\theta _{i}q^{t_{i}}}{%
\theta _{1}q^{t_{1}}}\beta _{1,i^{\prime
}}xG(x)\;\;\;\;\;\;\;\;\;\;\;\;\;\;\;\;\;\;\;\;\;i=2,\dots ,N-1  \notag \\
k_{i,1}(x) &=&\frac{\theta _{i}q^{t_{i}}}{\theta _{2}q^{t_{2}}}\frac{\beta
_{2,1}\beta _{1,i^{\prime }}}{\beta _{1,N-1}}G(x)\;\;\;\;\;\;\;\;\;\;\;\;\;%
\;\;\;\;\;\;\;\;\;\;\;\;i=3,\dots ,N-1  \notag \\
k_{N,j}(x) &=&-\frac{\epsilon }{\sqrt{\zeta }}\frac{\theta _{N}q^{t_{N}}}{%
\theta _{2}q^{t_{2}}}\frac{\beta _{2,1}\beta _{1,j}}{\beta _{1,N-1}}%
xG(x)\;\;\;\;\;\;\;\;\;\;\;\;\;\;j=2,\dots ,N-1  \notag \\
k_{1,j}(x) &=&\beta
_{1,j}G(x)\;\;\;\;\;\;\;\;\;\;\;\;\;\;\;\;\;\;\;\;\;\;\;\;\;\;\;\;\;\;\;\;\;%
\;\;\;\;\;\;\;\;\;\;\;\;\;j=2,\dots ,N-1.
\end{eqnarray}

The secondary diagonal is characterized by entries of the form 
\begin{equation}
k_{i,i^{\prime }}(x)=\left\{ 
\begin{array}{c}
\beta _{1,N}G(x)H_{f}(x),\qquad 
\hfill%
i=1 \\ 
\displaystyle%
-q^{2}\frac{\theta _{1}q^{t_{1}}}{\theta _{i^{\prime }}q^{t_{i^{\prime }}}}%
\left( \frac{1-\epsilon q\sqrt{\zeta }}{q+1}\right) ^{2}\frac{\beta
_{1,i^{\prime }}^{2}}{\beta _{1,N}}G(x)H_{f}(x),\qquad 
\hfill%
i\neq \{1,m+1,N\} \\ 
\displaystyle%
\frac{\theta _{N-1}q^{t_{N-1}}}{\theta _{2}q^{t_{2}}}\frac{\beta _{1,N}\beta
_{2,1}^{2}}{\beta _{1,N-1}^{2}}G(x)H_{f}(x),\qquad 
\hfill%
i=N.%
\end{array}%
\right.
\end{equation}%
and the remaining non-diagonal elements are given by 
\begin{equation}
k_{i,j}(x)=\left\{ 
\begin{array}{c}
\displaystyle%
-\frac{\epsilon }{\sqrt{\zeta }}\frac{\theta _{i}q^{t_{i}}}{\theta
_{1}q^{t_{1}}}\left( \frac{1-\epsilon q\sqrt{\zeta }}{q+1}\right) \frac{%
\beta _{1,i^{\prime }}\beta _{1,j}}{\beta _{1,N}}G(x),\qquad 
\hfill%
i<j^{\prime },\ 2<i,j<N-1 \\ 
\displaystyle%
\frac{1}{\zeta }\frac{\theta _{i}q^{t_{i}}}{\theta _{1}q^{t_{1}}}\left( 
\frac{1-\epsilon q\sqrt{\zeta }}{q+1}\right) \frac{\beta _{1,i^{\prime
}}\beta _{1,j}}{\beta _{1,N}}xG(x),\qquad 
\hfill%
i>j^{\prime },\ 2<i,j<N-1%
\end{array}%
\right. .
\end{equation}%
and%
\begin{eqnarray}
k_{1,m}(x) &=&\frac{i\sqrt{q}}{q-1}\frac{\beta _{1,m+1}^{2}}{\beta _{1,m+2}}%
G(x),  \notag \\
k_{2,1}(x) &=&(-1)^{m+1}q^{2m}\left( \frac{\epsilon q\sqrt{\zeta }-1}{q+1}%
\right) ^{2}\frac{\beta _{1,m}\beta _{1,m+2}\beta _{1,2m}}{\beta _{1,N}^{2}}%
G(x),  \notag \\
k_{1,N}(x) &=&%
\displaystyle%
\frac{i}{2}\frac{q(q^{\frac{3}{2}}+(-1)^{m}\epsilon )}{\sqrt{q}%
+(-1)^{m}\epsilon }\frac{(1+\epsilon \sqrt{\zeta })(\epsilon q\sqrt{\zeta }%
-1)}{\sqrt{\zeta }(q+1)^{2}}\beta _{1,m+1}^{2}G(x).
\end{eqnarray}

In their turn the diagonal entries $k_{\alpha ,\alpha }(x)$ are given by the
following expression%
\begin{eqnarray*}
k_{1,1}(x) &=&\left[ (\beta _{N,N}-\beta _{m+1,m+1})x-(\beta _{N,N}-\beta
_{1,1}-2)xH_{f}(x)+\beta _{m+1,m+1}-\beta _{1,1}\right] \frac{G(x)}{x^{2}-1}
\\
&&+\left( \frac{1+\epsilon \sqrt{\zeta }}{x^{2}-1}\right) \Delta (x),
\end{eqnarray*}%
with%
\begin{equation}
k_{i,i}(x)=\left\{ 
\begin{array}{c}
k_{1,1}(x)+(\beta _{i,i}-\beta _{1,1})G(x),\qquad 
\hfill%
i=2,...,m \\ 
k_{i-1,i-1}(x)+(\beta _{i,i}-\beta _{i-1,i-1})xG(x),\qquad 
\hfill%
i=m+3,...,N%
\end{array}%
\right.
\end{equation}%
and%
\begin{eqnarray*}
k_{m+1,m+1}(x) &=&k_{1,1}(x)+(\beta _{im+1,m+1}-\beta _{1,1})G(x)+\Gamma (x),
\\
k_{m+2,m+2}(x) &=&k_{m+1,m+1}(x)+(\beta _{im+2,m+2}-\beta
_{m+1,m+1})xG(x)+\epsilon \sqrt{\zeta }\Gamma (x),
\end{eqnarray*}%
where the auxiliary function $\Gamma (x)$ is given by 
\begin{equation}
\Gamma (x)=-\frac{q^{m+2}(\epsilon q\sqrt{\zeta }-1)}{(q+1)^{2}}\frac{\beta
_{1,m}\beta _{1,m+2}}{\beta _{1,N}}(x-1)G(x).
\end{equation}%
The parameters $\beta _{1,j}$ are constrained by the recurrence relation 
\begin{equation}
\beta _{1,j}=-\frac{\beta _{1,j+1}\beta _{1,N-j}}{\beta _{1,N+1-j}}%
,\;\;\;\;\;\;\;\;\;\;\;\;\;\;\;\;\;\;\;j=2,\dots ,m-1\;
\end{equation}%
while $\beta _{i,i}$ are fixed by 
\begin{equation}
\beta _{i,i}=\left\{ 
\begin{array}{c}
\displaystyle%
\beta _{1,1}+Q_{m}\sum_{k=0}^{i-2}(-\frac{1}{q})^{k},\qquad 
\hfill%
i=2,...,m \\ 
\beta _{m,m}+\Delta _{1},\qquad 
\hfill%
i=m+1 \\ 
\beta _{m,m}+\Delta _{2},\qquad 
\hfill%
i=m+2 \\ 
\displaystyle%
\beta _{m+2,m+2}-(-1)^{m}\epsilon \sqrt{q}Q_{m}\sum_{k=0}^{i-m-3}(-\frac{1}{q%
})^{k},\qquad 
\hfill%
i=m+3,...,N%
\end{array}%
\right.
\end{equation}%
where 
\begin{eqnarray}
Q_{m} &=&%
\displaystyle%
\frac{2(-1)^{m}q^{m-1}(q+1)^{2}\epsilon \sqrt{\zeta }}{(\sqrt{q}%
-(-1)^{m}\epsilon )(q^{\frac{3}{2}}+(-1)^{m}\epsilon )(1+\epsilon \sqrt{%
\zeta })},  \notag \\
\Delta _{1} &=&%
\displaystyle%
-\frac{2[q(1+\epsilon \sqrt{\zeta })+\epsilon \sqrt{\zeta }(q+1)]}{(\sqrt{q}%
-(-1)^{m}\epsilon )(q^{\frac{3}{2}}+(-1)^{m}\epsilon )(1+\epsilon \sqrt{%
\zeta })},  \notag \\
\Delta _{2} &=&%
\displaystyle%
\frac{2(q+1)\left( q-\epsilon \sqrt{\zeta }\right) }{(\sqrt{q}%
-(-1)^{m}\epsilon )(q^{\frac{3}{2}}+(-1)^{m}\epsilon )(1+\epsilon \sqrt{%
\zeta })}.
\end{eqnarray}%
In this solution the have a total amount of $m$ free parameters, namely $%
\beta _{1,m+1},\dots ,\beta _{1,2m}$.

\subsubsection{\textbf{Solution $\mathcal{M}_{10}$:}}

The series of solutions $\mathcal{M}_{10}$ is valid for the $%
U_{q}[osp(2|2m)^{(1)}]$ model and the corresponding $K$-matrix also possess
all entries different from zero. In the first and last columns, the matrix
elements are mainly given by 
\begin{eqnarray}
k_{i,1}(x) &=&\frac{\theta _{i}q^{t_{i}}}{\theta _{2}q^{t_{2}}}\frac{\beta
_{2,1}\beta _{1,i^{\prime }}}{\beta _{1,N-1}}G(x)\;\;\;\;\;\;\;\;\;\;\;\;\;%
\;\;\;\;\;\;i=3,\dots ,N-1  \notag \\
k_{i,N}(x) &=&-\frac{\epsilon }{\sqrt{\zeta }}\frac{\theta _{i}q^{t_{i}}}{%
\theta _{1}q^{tt_{1}}}\beta _{1,i^{\prime
}}xG(x)\;\;\;\;\;\;\;\;\;\;\;\;\;\;\;\;i=2,\dots ,N-1
\end{eqnarray}%
while the ones in the first and last rows are respectively 
\begin{eqnarray}
k_{1,j}(x) &=&\beta
_{1,j}G(x)\;\;\;\;\;\;\;\;\;\;\;\;\;\;\;\;\;\;\;\;\;\;\;\;\;\;\;\;\;\;\;\;\;%
\;\;\;\;\;\;\;\;\;\;\;j=2,\dots ,N-1  \notag \\
k_{N,j}(x) &=&-\frac{\epsilon }{\sqrt{\zeta }}\frac{\theta _{N}q^{t_{N}}}{%
\theta _{2}q^{t_{2}}}\frac{\beta _{2,1}\beta _{1,j}}{\beta _{1,N-1}}%
xG(x)\;\;\;\;\;\;\;\;\;\;\;\;j=2,\dots ,N-1
\end{eqnarray}

In the secondary diagonal we have the following expression determining the
matrix elements 
\begin{equation}
k_{i,i^{\prime }}(x)=\left\{ 
\begin{array}{c}
\beta _{1,N}G(x)H_{f}(x),\qquad i=1 \\ 
\displaystyle%
-q^{2}\frac{\theta _{1}q^{t_{1}}}{\theta _{i^{\prime }}q^{t_{i^{\prime }}}}%
\left( \frac{1-\epsilon q\sqrt{\zeta }}{q+1}\right) ^{2}\frac{\beta
_{1,i^{\prime }}^{2}}{\beta _{1,N}}G(x)H_{f}(x),\qquad i\neq \{1,m+1,m+2,N\}
\\ 
\displaystyle%
-q^{2}\frac{\theta _{1}q^{t_{1}}}{\theta _{i^{\prime }}q^{t_{i^{\prime }}}}%
\frac{(q+\epsilon \sqrt{\zeta })(1-\epsilon q\sqrt{\zeta })}{q^{2}-1}\frac{%
\beta _{1,i^{\prime }}^{2}}{\beta _{1,N}}G(x)H_{b}(x),\qquad i=\{m+1,m+2\}
\\ 
\displaystyle%
\frac{\theta _{N-1}q^{t_{N-1}}}{\theta _{2}q^{t_{2}}}\frac{\beta _{1,N}\beta
_{2,1}^{2}}{\beta _{1,N-1}^{2}}G(x)H_{f}(x),\qquad i=N.%
\end{array}%
\right.
\end{equation}%
recalling that 
\begin{equation}
H_{b}(x)=\frac{qx+\epsilon \sqrt{\zeta }}{q+\epsilon \sqrt{\zeta }}%
\;\;\;\;\;\;H_{f}(x)=\frac{x-\epsilon q\sqrt{\zeta }}{1-\epsilon q\sqrt{%
\zeta }}.  \label{hfhb}
\end{equation}

In their turn the other non-diagonal entries satisfy the relation 
\begin{equation}
k_{i,j}(x)=\left\{ 
\begin{array}{c}
\displaystyle%
-\frac{\epsilon }{\sqrt{\zeta }}\frac{\theta _{i}q^{t_{i}}}{\theta
_{1}q^{t_{1}}}\left( \frac{1-\epsilon q\sqrt{\zeta }}{q+1}\right) \frac{%
\beta _{1,i^{\prime }}\beta _{1,j}}{\beta _{1,N}}G(x),\qquad i<j^{\prime },\
2<i,j<N-1 \\ 
\displaystyle%
\frac{1}{\zeta }\frac{\theta _{i}q^{t_{i}}}{\theta _{1}q^{t_{1}}}\left( 
\frac{1-\epsilon q\sqrt{\zeta }}{q+1}\right) \frac{\beta _{1,i^{\prime
}}\beta _{1,j}}{\beta _{1,N}}xG(x),\qquad i>j^{\prime },\ 2<i,j<N-1%
\end{array}%
\right. ,
\end{equation}%
and%
\begin{eqnarray}
k_{1,m}(x) &=&\frac{i(q+1)}{q-1}\frac{\beta _{1,m+1}\beta _{1,m+2}}{\beta
_{1,m+3}}G(x),  \notag \\
k_{1,m+1}(x) &=&%
\displaystyle%
\frac{2i\epsilon \sqrt{\zeta }(q+1)}{q(\sqrt{\zeta }-(-1)^{m})(q\sqrt{\zeta }%
+(-1)^{m})((-1)^{m}-\epsilon )}\frac{\beta _{1,N}}{\beta _{1,m+2}}G(x), 
\notag \\
k_{2,1}(x) &=&\frac{i(-1)^{m+1}}{q\zeta }\left( \frac{q\sqrt{\zeta }%
-\epsilon }{q\sqrt{\zeta }+\epsilon }\right) \left( \frac{q^{2}\zeta -1}{%
q^{2}-1}\right) \frac{\beta _{1,m+1}\beta _{1,m+2}\beta _{1,N-1}}{\beta
_{1,N}^{2}}G(x),
\end{eqnarray}%
and the parameters $\beta _{1,j}$ are required to satisfy the recurrence
relation 
\begin{equation}
\beta _{1,j}=-\frac{\beta _{1,j+1}\beta _{1,N-j}}{\beta _{1,N-j+1}}%
\;\;\;\;\;\;\;\;\;\;\;\;\;\;\;j=2,\dots ,m-1.
\end{equation}

Considering now the diagonal entries, they are given by%
\begin{eqnarray}
k_{1,1}(x) &=&\left[ (\beta _{N,N}-\beta _{m+1,m+1})x-(\beta _{N,N}-\beta
_{1,1}-2)xH_{f}(x)+\beta _{m+1,m+1}-\beta _{1,1}\right] \frac{G(x)}{x^{2}-1},
\notag \\
k_{i,i}(x) &=&\left\{ 
\begin{array}{c}
k_{1,1}(x)+(\beta _{i,i}-\beta _{1,1})G(x),\qquad 
\hfill%
i=2,...,m+1 \\ 
k_{m+1,m+1}(x),\qquad 
\hfill%
i=m+2 \\ 
k_{m+1,m+1}(x)+(\beta _{i,i}-\beta _{m+1,m+1})xG(x),\qquad 
\hfill%
i=m+3,...,N%
\end{array}%
\right. ,
\end{eqnarray}%
where the parameters $\beta _{\alpha ,\alpha }$ are determined by the
expressions 
\begin{equation}
\beta _{i,i}=\left\{ 
\begin{array}{c}
\displaystyle%
\beta _{1,1}+\Delta _{m}\sum_{k=0}^{i-2}(-\frac{1}{q})^{k},\qquad 
\hfill%
i=2,...,m \\ 
\displaystyle%
\beta _{m,m}+\frac{4\sqrt{\zeta }(1-\epsilon \sqrt{\zeta })}{(q-1)(\sqrt{%
\zeta }-(-1)^{m})(q\sqrt{\zeta }+(-1)^{m})(\epsilon -(-1)^{m})},\qquad 
\hfill%
i=m+1,m+2 \\ 
\displaystyle%
\beta _{m+1,m+1}+\frac{[q^{j-m-2}-q^{j-m-3}-(-1)^{j-m}(q+1)]q^{N-j+1}}{%
(q-1)(q^{m}-(-1)^{m})},\qquad 
\hfill%
i=m+3,...,N%
\end{array}%
\right.
\end{equation}%
with 
\begin{equation}
\Delta _{m}=\frac{2(-1)^{m}(q+1)^{2}}{q^{2}(q-1)(1-(-1)^{m}\sqrt{\zeta }%
)(\epsilon -(-1)^{m})}.
\end{equation}%
This solution has altogether $m+1$ free parameters, namely $\beta
_{1,m+2},\dots ,\beta _{1,N}$.

\subsubsection{\textbf{Solution $\mathcal{M}_{11}$:}}

The class of solutions $\mathcal{M}_{11}$ is valid for the vertex model $%
U_{q}[osp(2n|2)^{(1)}]$ and the corresponding $K$-matrix contains only
non-null entries. The border elements are mainly given by the following
expressions 
\begin{eqnarray}
k_{i,N}(x) &=&-\frac{\epsilon }{\sqrt{\zeta }}\frac{\theta _{i}q^{t_{i}}}{%
\theta _{1}q^{t_{1}}}\beta _{1,i^{\prime
}}xG(x)\;\;\;\;\;\;\;\;\;\;\;\;\;\;\;\;\;\;\;\;\;i=2,\dots ,N-1  \notag \\
k_{i,1}(x) &=&\frac{\theta _{i}q^{t_{i}}}{\theta _{2}q^{t_{2}}}\frac{\beta
_{2,1}\beta _{1,i^{\prime }}}{\beta _{1,N-1}}G(x)\;\;\;\;\;\;\;\;\;\;\;\;\;%
\;\;\;\;\;\;\;\;\;\;\;\;i=3,\dots ,N-1  \notag \\
k_{N,j}(x) &=&-\frac{\epsilon }{\sqrt{\zeta }}\frac{\theta _{N}q^{t_{N}}}{%
\theta _{2}q^{t_{2}}}\frac{\beta _{2,1}\beta _{1,j}}{\beta _{1,N-1}}%
xG(x)\;\;\;\;\;\;\;\;\;\;\;\;\;\;j=2,\dots ,N-1  \notag \\
k_{1,j}(x) &=&\beta
_{1,j}G(x)\;\;\;\;\;\;\;\;\;\;\;\;\;\;\;\;\;\;\;\;\;\;\;\;\;\;\;\;\;\;\;\;\;%
\;\;\;\;\;\;\;\;\;\;\;\;\;j=2,\dots ,N-1.
\end{eqnarray}

The secondary diagonal is constituted by elements $k_{i,i^{\prime }}(x)$
given by%
\begin{equation}
k_{i,i^{\prime }}(x)=\left\{ 
\begin{array}{c}
\beta _{1,N}G(x)H_{f}(x),\qquad i=1 \\ 
\displaystyle%
-q^{2}\frac{\theta _{1}q^{t_{1}}}{\theta _{i^{\prime }}q^{t_{i^{\prime }}}}%
\left( \frac{1-\epsilon q\sqrt{\zeta }}{q-1}\right) \left( \frac{q+\epsilon 
\sqrt{\zeta }}{q+1}\right) \frac{\beta _{1,i^{\prime }}^{2}}{\beta _{1,N}}%
G(x)H_{b}(x),\qquad i\neq \{1,N\} \\ 
\displaystyle%
\frac{\theta _{N-1}q^{t_{N-1}}}{\theta _{2}q^{t_{2}}}\frac{\beta _{1,N}\beta
_{2,1}^{2}}{\beta _{1,N-1}^{2}}G(x)H_{f}(x),\qquad i=N.%
\end{array}%
\right.
\end{equation}%
where the functions $H_{b}(x)$ and $H_{f}(x)$ were already given in (\ref%
{hfhb}). The remaining non-diagonal entries are determined by the expression 
\begin{equation}
k_{i,j}(x)=\left\{ 
\begin{array}{c}
\displaystyle%
\frac{\epsilon }{\sqrt{\zeta }}\frac{\theta _{i}q^{t_{i}}}{\theta
_{1}q^{t_{1}}}\left( \frac{1-\epsilon q\sqrt{\zeta }}{q-1}\right) \frac{%
\beta _{1,i^{\prime }}\beta _{1,j}}{\beta _{1,N}}G(x),\qquad i<j^{\prime },\
2<i,j<N-1 \\ 
\displaystyle%
\frac{1}{\zeta }\frac{\theta _{i}q^{t_{i}}}{\theta _{1}q^{t_{1}}}\left( 
\frac{1-\epsilon q\sqrt{\zeta }}{q-1}\right) \frac{\beta _{1,i^{\prime
}}\beta _{1,j}}{\beta _{1,N}}xG(x),\qquad i>j^{\prime },\ 2<i,j<N-1%
\end{array}%
\right. ,
\end{equation}%
and%
\begin{equation*}
k_{1,n+1}(x)=%
\displaystyle%
\frac{i\sqrt{\zeta }(q-1)}{q(q^{n-1}+(-1)^{n})(q^{n-2}+(-1)^{n})}\frac{\beta
_{1,N}}{\beta _{1,n+2}}G(x)
\end{equation*}%
\begin{equation*}
k_{2,1}(x)=\frac{i(-1)^{n}}{(q+1)q^{n-1}}\left( \frac{q^{n-1}+(-1)^{n}}{%
q^{n-2}+(-1)^{n}}\right) \frac{\beta _{1,N-1}}{\beta _{1,N}}G(x).
\end{equation*}%
and the parameters $\beta _{1,j}$ are required to satisfy 
\begin{equation}
\beta _{1,j}=-\frac{\beta _{1,j+1}\beta _{1,N-j}}{\beta _{1,N+1-j}}%
\;\;\;\;\;\;\;\;\;\;\;\;\;\;\;\;\;\;\;\;\;j=2,\dots ,n.
\end{equation}

With respect to the diagonal entries, they are given by%
\begin{eqnarray}
k_{1,1}(x) &=&\left[ (\beta _{N,N}-\beta _{n+1,n+1})x-(\beta _{N,N}-\beta
_{1,1}-2)xH_{f}(x)+\beta _{n+1,n+1}-\beta _{1,1}\right] \frac{G(x)}{x^{2}-1},
\notag \\
k_{i,i}(x) &=&\left\{ 
\begin{array}{c}
k_{1,1}(x)+(\beta _{i,i}-\beta _{1,1})G(x),\qquad 
\hfill%
i=2,...,n+1 \\ 
k_{n+1,n+1}(x),\qquad 
\hfill%
i=n+2 \\ 
k_{n+1,n+1}(x)+(\beta _{i,i}-\beta _{n+1,n+1})xG(x),\qquad 
\hfill%
i=n+3,...,N%
\end{array}%
\right. ,
\end{eqnarray}%
where the parameters $\beta _{\alpha ,\alpha }$ are determined by the
expressions 
\begin{equation}
\beta _{i,i}=\left\{ 
\begin{array}{c}
\displaystyle%
\beta _{1,1}-\frac{(-1)^{n+i}}{q(q+1)(q^{n-2}+(-1)^{n})}%
[q^{i-1}+q^{i-2}-(-1)^{i}(q-1)],\qquad 
\hfill%
i=2,...,n+1 \\ 
\displaystyle%
\beta _{n+1,n+1},\qquad 
\hfill%
i=n+2 \\ 
\displaystyle%
\beta _{n+1,n+1}+\frac{(-1)^{n}(q+1)}{(q^{n-2}+(-1)^{n})}%
\sum_{k=0}^{i-n-3}(-q)^{k},\qquad 
\hfill%
i=n+3,...,N-1 \\ 
\beta _{1,1}+2-\frac{q^{2}+1}{q(q+1)}\left( \frac{q^{n-1}+(-1)^{n}}{%
q^{n-2}+(-1)^{n}}\right) .\qquad 
\hfill%
i=N%
\end{array}%
\right.
\end{equation}%
The variables $\beta _{1,n+2},\dots ,\beta _{1,N}$ give us a total amount of 
$n+1$ free parameters.

\subsubsection{\textbf{Solution $\mathcal{M}_{12}$:}}

The solution $\mathcal{M}_{12}$ also does not contain null entries and it is
valid for the $U_{q}[osp(2n+1|2)^{(1)}]$ vertex model. Considering first the
non-diagonal entries, we have the following expression determining border
elements, 
\begin{eqnarray}
k_{i,N}(x) &=&-\frac{\epsilon }{\sqrt{\zeta }}\frac{\theta _{i}q^{t_{i}}}{%
\theta _{1}q^{t_{1}}}\beta _{1,i^{\prime
}}xG(x)\;\;\;\;\;\;\;\;\;\;\;\;\;\;\;\;\;\;\;\;\;\;\;\;\;\;\;\;\;\;\;\;\;\;%
\;i=2,\dots ,N-1  \notag \\
k_{i,1}(x) &=&\frac{\theta _{i}q^{t_{i}}}{\theta _{2}q^{t_{2}}}\frac{\beta
_{2,1}\beta _{1,i^{\prime }}}{\beta _{1,N-1}}G(x)\;\;\;\;\;\;\;\;\;\;\;\;\;%
\;\;\;\;\;\;\;\;\;\;\;\;\;\;\;\;\;\;\;\;\;\;\;\;\;\;i=3,\dots ,N-1  \notag \\
k_{N,j}(x) &=&-\frac{\epsilon }{\sqrt{\zeta }}\frac{\theta _{N}q^{t_{N}}}{%
\theta _{2}q^{t_{2}}}\frac{\beta _{2,1}\beta _{1,j}}{\beta _{1,N-1}}%
xG(x)\;\;\;\;\;\;\;\;\;\;\;\;\;\;\;\;\;\;\;\;\;\;\;\;\;\;\;\;\;j=2,\dots ,N-1
\notag \\
k_{1,j}(x) &=&\beta
_{1,j}G(x)\;\;\;\;\;\;\;\;\;\;\;\;\;\;\;\;\;\;\;\;\;\;\;\;\;\;\;\;\;\;\;\;\;%
\;\;\;\;\;\;\;\;\;\;\;\;\;\;\;\;\;\;\;\;\;\;\;\;\;\;\;\;j=2,\dots ,N-1,
\end{eqnarray}%
and the following one for the entries of the secondary diagonal%
\begin{equation}
k_{i,i^{\prime }}(x)=\left\{ 
\begin{array}{c}
\beta _{1,N}G(x)H_{f}(x),\qquad i=1 \\ 
\displaystyle%
-q^{2}\frac{\theta _{1}q^{t_{1}}}{\theta _{i^{\prime }}q^{t_{i^{\prime }}}}%
\left( \frac{1-\epsilon q\sqrt{\zeta }}{q-1}\right) \left( \frac{q+\epsilon 
\sqrt{\zeta }}{q+1}\right) \frac{\beta _{1,i^{\prime }}^{2}}{\beta _{1,N}}%
G(x)H_{b}(x),\qquad i\neq \{1,i^{\prime },N\} \\ 
\displaystyle%
\frac{\theta _{N-1}q^{t_{N-1}}}{\theta _{2}q^{t_{2}}}\frac{\beta _{1,N}\beta
_{2,1}^{2}}{\beta _{1,N-1}^{2}}G(x)H_{f}(x),\qquad i=N.%
\end{array}%
\right.
\end{equation}%
where the functions $H_{b}(x)$ and $H_{f}(x)$ were already given in (\ref%
{hfhb}). The remaining non-diagonal entries are determined by the expression 
\begin{equation}
k_{i,j}(x)=\left\{ 
\begin{array}{c}
\displaystyle%
\frac{\epsilon }{\sqrt{\zeta }}\frac{\theta _{i}q^{t_{i}}}{\theta
_{1}q^{t_{1}}}\left( \frac{1-\epsilon q\sqrt{\zeta }}{q-1}\right) \frac{%
\beta _{1,i^{\prime }}\beta _{1,j}}{\beta _{1,N}}G(x),\qquad i<j^{\prime },\
2<i,j<N-1 \\ 
\displaystyle%
\frac{1}{\zeta }\frac{\theta _{i}q^{t_{i}}}{\theta _{1}q^{t_{1}}}\left( 
\frac{1-\epsilon q\sqrt{\zeta }}{q-1}\right) \frac{\beta _{1,i^{\prime
}}\beta _{1,j}}{\beta _{1,N}}xG(x),\qquad i>j^{\prime },\ 2<i,j<N-1%
\end{array}%
\right. ,
\end{equation}%
and%
\begin{eqnarray}
k_{2,1}(x) &=&\frac{(-1)^{n}}{\zeta }\frac{(\epsilon q\sqrt{\zeta }-1)^{2}}{%
q^{2}-1}\frac{\beta _{1,n+1}\beta _{1,n+3}\beta _{1,N-1}}{\beta _{1,N}^{2}}%
G(x),  \notag \\
k_{1,n+1}(x) &=&%
\displaystyle%
\frac{\epsilon (-1)^{n}(q+1)}{(\sqrt{q}-(-1)^{n}\epsilon )^{2}}\left[ \frac{%
\beta _{1,n+2}^{2}}{\beta _{1,n+3}}-\frac{2i(q-1)\sqrt{\zeta }}{q(q\sqrt{%
\zeta }-\epsilon )(\sqrt{\zeta }-\epsilon )}\frac{\beta _{1,N}}{\beta
_{1,n+2}}\right] G(x),
\end{eqnarray}%
and the parameters $\beta _{1,j}$ are required to satisfy 
\begin{equation}
\beta _{1,j}=-\frac{\beta _{1,j+1}\beta _{1,N-j}}{\beta _{1,N+1-j}}%
\;\;\;\;\;\;\;\;\;\;\;\;\;\;\;\;\;\;\;\;\;j=2,\dots ,n.
\end{equation}

With respect to the diagonal entries, they are given by%
\begin{eqnarray}
k_{1,1}(x) &=&\left[ (\beta _{N,N}-\beta _{n+2,n+2})x-(\beta _{N,N}-\beta
_{1,1}-2)xH_{f}(x)+\beta _{n+2,n+2}-\beta _{1,1}\right] \frac{G(x)}{x^{2}-1}
\notag \\
&&+\left( \frac{1+\epsilon \sqrt{\zeta }}{x^{2}-1}\right) \Delta (x) \\
k_{i,i}(x) &=&\left\{ 
\begin{array}{c}
k_{1,1}(x)+(\beta _{i,i}-\beta _{1,1})G(x),\qquad i=2,...,n+1 \\ 
k_{n+1,n+1}(x)+(\beta _{n+2,n+2}-\beta _{n+1,n+1})G(x)+\Delta (x),\qquad
i=n+2 \\ 
k_{n+2,n+2}(x)+(\beta _{n+3,n+3}-\beta _{n+2,n+2})xG(x)+\epsilon \sqrt{\zeta 
}\Delta (x),\qquad i=n+3 \\ 
k_{n+3,n+3}(x)+(\beta _{i,i}-\beta _{n+3,n+3})xG(x),\qquad i=n+4,...,N%
\end{array}%
\right.  \notag
\end{eqnarray}%
where%
\begin{equation}
\Delta (x)=\frac{i(\epsilon q\sqrt{\zeta }-1)}{q^{n-3}(q^{2}-1)}\frac{\beta
_{1,n+1}\beta _{1,n+3}}{\beta _{1,N}}(x-1)G(x)
\end{equation}%
The parameters $\beta _{\alpha ,\alpha }$ are determined by the expressions 
\begin{equation}
\beta _{i,i}=\left\{ 
\begin{array}{c}
\displaystyle%
\beta _{1,1}+(-1)^{n}Q_{n}[(-q)^{i-2}(q+1)+(1-q)],\qquad i=2,...,n+1 \\ 
\displaystyle%
\beta _{n+1,n+1}-\frac{iq(q\sqrt{\zeta }-\epsilon )}{q-1}\left[ \frac{\beta
_{1,n+2}^{2}}{\beta _{1,N}}-\frac{(\epsilon q+\sqrt{\zeta })}{q^{n-1}(q+1)}%
\frac{\beta _{1,n+1}\beta _{1,n+3}}{\beta _{1,N}}\right] ,\qquad i=n+2 \\ 
\displaystyle%
\beta _{n+2,n+2}+\frac{iq(q\sqrt{\zeta }-\epsilon )}{(q-1)}\left[ \frac{%
\epsilon \beta _{1,n+2}^{2}}{\sqrt{\zeta }\beta _{1,N}}-\frac{(\epsilon q+%
\sqrt{\zeta })}{q^{n-2}(q+1)}\frac{\beta _{1,n+1}\beta _{1,n+3}}{\beta _{1,N}%
}\right] ,\qquad i=n+3 \\ 
\displaystyle%
\beta _{n+3,n+3}+\frac{(q+1)^{2}}{q^{n-3}}\epsilon \sqrt{\zeta }%
Q_{n}\sum_{k=0}^{i-n-4}(-q)^{k},\qquad i=n+4,...,N-1 \\ 
\displaystyle%
\beta _{1,1}+2-\frac{i\epsilon (-1)^{n}}{\sqrt{\zeta }}\left( \frac{q^{2}+1}{%
q^{2}-1}\right) (\epsilon q\sqrt{\zeta }-1)^{2}\frac{\beta _{1,n+1}\beta
_{1,n+3}}{\beta _{1,N}},\qquad i=N%
\end{array}%
\right.
\end{equation}%
and the auxiliary parameter $Q_{n}$ is given by 
\begin{equation}
Q_{n}=\frac{i\epsilon (\epsilon q\sqrt{\zeta }-1)}{\sqrt{\zeta }(q^{2}-1)}%
\frac{\beta _{1,n+1}\beta _{1,n+3}}{\beta _{1,N}}.
\end{equation}%
This solution possess $n+2$ free parameters, namely $\beta _{1,n+2},\dots
,\beta _{1,N}$.

\section{Concluding Remarks}

In this work we have presented the general set of regular solutions of the
graded reflection equation for the $U_{q}[osp(r|2m)^{(1)}]$ vertex model.
Our findings can be summarized into four classes of diagonal solutions and
twelve classes of non-diagonal ones. Although the $R$ matrix of the $%
U_{q}[sl(r|2m)^{(2)}]$ vertex model is similar to the $R$ matrix of the $%
U_{q}[osp(r|2m)^{(1)}]$ vertex model, their $K$-matrices solutions are
different by the number of free parameters in each solution (see \cite{LIMG}%
).

It seems too much \ the number of the $K$-matrix solutions. We have tried to
use a particular gauge transformation in order to relate them but without
success. The main difficult is the difference of the number of free
parameters in each class of solution $\mathcal{M}_{i}$.

Here we remark that we know from the $osp(1|2)^{(1)}$ and $sl(2|1)^{(2)}$
models that solutions of the graded Yang-Baxter equation can be mapped to
(the $A_{2}^{(2)}$ and $B_{1}^{(1)}$ models, respectively) solutions of the
non-graded equations by graded permutations and gauge transformations \cite%
{Martins, Fireman}. \ 

Although our complete solutions were derived from solutions based on
superalgebras, we could use the non-graded formalism in order to construct
them. Thus, the $K$ matrices with both bosonic and fermionic entries are
indicating a symmetry breaking at the boundaries but do not have any
contradiction with the corresponding boundary integrability.

These results pave the way to construct, solve and study physical properties
of the underlying quantum spin chains with open boundaries, generalizing the
previous efforts made for the case of periodic boundary conditions \cite%
{GALLEAS1, GALLEAS2}.

Although we expect that the Algebraic Bethe Ansatz solution of the models
constructed from the diagonal solutions presented here can be obtained by
adapting the results of \cite{G4}, the algebraic-functional method presented
in \cite{GALLEAS4} may be a possibility to treat the non-diagonal cases.

For further research, an interesting possibility would be the investigation
of soliton non-preserving boundary conditions \cite{G3,DO1} for quantum spin
chains based on $q$-deformed Lie algebras and superalgebras, which can also
be performed by adapting the method described in \cite{LIM}. We expect the
results presented here to motivate further developments on the subject of
integrable open boundaries for vertex models based on $q$-deformed Lie
superalgebras. In particular, the classification of the solutions of the
graded reflection equation for others $q$-deformed Lie superalgebras, which
we hope to report on a future work.

\section{Acknowledgments}

We thank to W. Galleas for his valuable discussions. This work is partially
supported by the Brazilian research councils CNPq and FAPESP.

\section*{\textbf{Appendix A: The $U_{q}[osp(1|2)^{(1)}]$ case}}

\setcounter{equation}{0} \renewcommand{\theequation}{A.\arabic{equation}}

The reflection equation associated with the $U_{q}[osp(1|2)^{(1)}]$ vertex
model admits more general solutions than the corresponding ones obtained
from the general series presented in the section 3. In this case the
reflection matrices were previously studied in \cite{LIM1} and we have
obtained the following solutions 
\begin{equation}
K^{-}(x)=\mathrm{Diag}(1,\frac{q^{\frac{3}{2}}+\epsilon x}{q^{\frac{3}{2}%
}+\epsilon },x^{2})  \label{a12}
\end{equation}%
and%
\begin{equation}
K^{-}(x)=\left( 
\begin{array}{ccc}
\displaystyle%
x-\frac{1}{2}\frac{\beta (q+1)}{q}(x-1) & 0 & 
\displaystyle%
\frac{1}{2}\beta _{13}(x^{2}-1) \\ 
0 & 
\displaystyle%
x+\frac{1}{2}\frac{\beta }{q}(xq-1)(x-1) & 0 \\ 
\displaystyle%
-\frac{1}{2}\frac{\beta ^{2}}{q\beta _{13}}(x^{2}-1) & 0 & 
\displaystyle%
x+\frac{1}{2}\frac{\beta (q+1)}{q}x(x-1)%
\end{array}%
\right)  \label{b12}
\end{equation}%
\noindent where we have two free parameters $\beta =\beta _{22}-\beta _{11}$
and $\beta _{13}$.

In addition to the solutions (\ref{a12}) and (\ref{b12}) we also have a
solution in the general form 
\begin{equation}
K^{-}(x)=\left( 
\begin{array}{ccc}
k_{11}(x) & \beta _{12}G(x) & 
\displaystyle%
\beta _{13}\frac{x\sqrt{q}-\epsilon }{\sqrt{q}-\epsilon }G(x) \\ 
\beta _{21}G(x) & k_{22}(x) & -i\epsilon q\beta _{12}xG(x) \\ 
\displaystyle%
\beta _{13}\left( \frac{\beta _{21}}{\beta _{12}}\right) ^{2}\frac{x\sqrt{q}%
-\epsilon }{\sqrt{q}-\epsilon }G(x) & -i\epsilon q\beta _{21}xG(x) & 
k_{33}(x)%
\end{array}%
\right) .
\end{equation}%
where%
\begin{equation}
\beta _{21}=\frac{\epsilon q^{\frac{3}{2}}}{(q-1)}\left[ \frac{\beta
_{12}^{2}}{\beta _{13}}-\frac{2i(\sqrt{q}+\epsilon )}{\sqrt{q}(q^{\frac{3}{2}%
}-\epsilon )}\right] \frac{\beta _{12}}{\beta _{13}}
\end{equation}%
The diagonal entries are then given by 
\begin{eqnarray}
k_{11}(x) &=&\frac{(x\sqrt{q}-\epsilon )}{(\sqrt{q}-\epsilon )^{2}}\left[ 
\frac{x(1+q^{2})-\epsilon \sqrt{q}(q+1)}{q^{\frac{3}{2}}-\epsilon }\right] 
\frac{2G(x)}{x^{2}-1}  \notag \\
&&+\left[ \frac{x\sqrt{q}(1+q^{2})-\sqrt{q}(q-1)-\epsilon (q+1)}{(\sqrt{q}%
-\epsilon )(q-1)}\right] \frac{i\sqrt{q}\beta _{12}^{2}}{\beta _{13}}\frac{%
G(x)}{x+1}  \notag \\
k_{22}(x) &=&k_{11}(x)+(\beta _{22}-\beta _{11})G(x)+\Delta (x)  \notag \\
k_{33}(x) &=&x^{2}k_{11}(x)+(\beta _{33}-\beta _{11}-2)\left( \frac{x\sqrt{q}%
-\epsilon }{\sqrt{q}-\epsilon }\right) xG(x),
\end{eqnarray}%
where 
\begin{equation}
\Delta (x)=-\frac{\epsilon q^{2}}{(\sqrt{q}-\epsilon )}\left[ \frac{i\sqrt{q}%
}{(q-1)}\frac{\beta _{12}^{2}}{\beta _{13}}+\frac{2}{(\sqrt{q}-\epsilon )(q^{%
\frac{3}{2}}-\epsilon )}\right] (x-1)G(x).
\end{equation}%
This solution has altogether two free parameters $\beta _{1,2}$ and $\beta
_{1,3}$ and the remaining variables $\beta _{i,j}$ are given by 
\begin{eqnarray}
\beta _{22} &=&\beta _{11}-\frac{2\epsilon q^{\frac{3}{2}}}{(\sqrt{q}%
-\epsilon )(q^{\frac{3}{2}}-\epsilon )}-\left( \frac{1}{q-1}+\epsilon \sqrt{q%
}\right) \frac{i\sqrt{q}\beta _{12}^{2}}{\beta _{13}},  \notag \\
\beta _{33} &=&\beta _{11}-\frac{2\epsilon \sqrt{q}(q+1)}{(\sqrt{q}-\epsilon
)(q^{\frac{3}{2}}-\epsilon )}-\left( \frac{1+q^{2}}{q-1}\right) \frac{i\sqrt{%
q}\beta _{12}^{2}}{\beta _{13}}.
\end{eqnarray}

\section*{\textbf{Appendix B: The $U_{q}[osp(2|2)^{(1)}]$ case}}

\setcounter{equation}{0} \renewcommand{\theequation}{B.\arabic{equation}}
The set of $K$-matrices associated with the $U_{q}[osp(2|2)^{(1)}]$ vertex
model includes both diagonal and non-diagonal solutions. The three solutions
intrinsically diagonal two contain only one free parameter $\beta $ and they
are given by 
\begin{eqnarray}
K^{-}(x) &=&\mathrm{Diag}(1,x\frac{\beta (x-1)+2}{\beta (x-1)-2x},x\frac{%
\beta (q^{4}-x)+2x}{\beta (xq^{4}-1)+2},x^{2}),  \notag \\
K^{-}(x) &=&\mathrm{Diag}(1,1,x\frac{\beta (x-1)+2}{\beta (x-1)-2x},x\frac{%
\beta (x-1)+2}{\beta (x-1)-2x})
\end{eqnarray}%
and one without free parameters 
\begin{equation}
K^{-}(x)=\mathrm{Diag}(1,x\frac{q^{2}+\epsilon x}{xq^{2}+\epsilon },x\frac{%
q^{2}+\epsilon x}{xq^{2}+\epsilon },x^{2}).
\end{equation}%
We have also found the following non-diagonal solutions 
\begin{equation}
K^{-}(x)=\left( 
\begin{array}{cccc}
\displaystyle%
x+\frac{\alpha -\beta ^{2}}{2\beta }(x-1) & 0 & 0 & 
\displaystyle%
\frac{\beta _{14}}{2}(x^{2}-1) \\ 
0 & 
\displaystyle%
x+\frac{\alpha +x\beta ^{2}}{2\beta }(x-1) & 0 & 0 \\ 
0 & 0 & 
\displaystyle%
x+\frac{\alpha +x\beta ^{2}}{2\beta }(x-1) & 0 \\ 
\displaystyle%
\frac{\alpha }{2\beta _{14}}(x^{2}-1) & 0 & 0 & 
\displaystyle%
x-\frac{\alpha -\beta ^{2}}{2\beta }x(x-1)%
\end{array}%
\right)
\end{equation}%
containing three free parameters $\alpha ,\beta $ and $\beta _{14}$, and one
free parameter solution in the form 
\begin{equation}
K^{-}(x)=\left( 
\begin{array}{cccc}
\displaystyle%
\left( \frac{q^{2}x^{2}-1}{q^{2}-1}\right) \frac{1}{x} & 0 & 0 & 0 \\ 
0 & x & 
\displaystyle%
\frac{1}{2}\beta _{23}(x^{2}-1) & 0 \\ 
0 & 
\displaystyle%
\frac{2}{\beta _{23}}\left( \frac{q}{q^{2}-1}\right) ^{2}(x^{2}-1) & x & 0
\\ 
0 & 0 & 0 & 
\displaystyle%
\left( \frac{q^{2}x^{2}-1}{q^{2}-1}\right) x%
\end{array}%
\right) .  \label{b5}
\end{equation}%
Concerning the complete solution, we could not find a complete solution for
this model.

\newpage%

\section*{\textbf{Appendix C: The main reflection equations}}

\setcounter{equation}{0} \renewcommand{\theequation}{C.\arabic{equation}}

In this appendix we present the main sequence of the step necessary to get a
general solution of the reflection equation. First we consider the (i,j)
component of the matrix equation (\ref{RE}). By differentiating it with
respect to $v$ and by taking $v=0$, we obtain $N^{4}$ algebraic equations $%
E[i,j]=0$ involving the single variable $u$ and $N^{2}$ parameters $\beta
_{i,j}$ for the matrix elements $k_{i,j}(u)$. \ Although there are many
equations ($N=r+2m,\quad r\geq 1,\quad m\geq 1$ ) a few of them are actually
independent.

Analyzing these $E[i,j]=0$ equations, the simplest are those involving only
two matrix elements of the type $k_{i,i^{\prime }}(u)$ (secondary diagonal)%
\begin{equation}
k_{i,i^{\prime }}(u)=\left\{ 
\begin{array}{c}
\frac{\beta _{i,i^{\prime }}}{\beta _{1,N}}k_{1,N}(u),\qquad 
\hfill%
i=1,2,...,m \\ 
\frac{\beta _{i,i^{\prime }}}{\beta _{m+1,N-m}}k_{m+1,N-m}(u),\qquad
i=m+1,...,m+r \\ 
\frac{\beta _{i,i^{\prime }}}{\beta _{1,N}}k_{1,N}(u),\qquad 
\hfill%
i=m+r+1,...,N%
\end{array}%
\right. .  \label{c1}
\end{equation}%
Moreover, we can use the following pairs of equations%
\begin{equation}
E[i+iN-N,N^{2}-iN+j]=0\quad \mathrm{and\quad }%
E[N^{2}+1-(N^{2}-iN+j),N^{2}+1-(i+iN-N)]=0  \label{c2}
\end{equation}%
for each $i=\{1,2,\cdots ,m\}$ with\textrm{\ \ }$j=i+1,\cdots ,N+1-i$, in
order to get the matrix elements $k_{i,j}(u)$ and $k_{N+1-j,N+1-i}(u)$ in
terms of $k_{i,i^{\prime }}(u)$.

This procedure allows to find all non-diagonal entries $k_{i,j}(u)$ with
fermionic degree of freedom ($i,j=1,2,...,m$ and\ $m+r+1,...,N$ ) in terms
of $k_{1,N}(u)$ 
\begin{equation}
k_{i,j}(u)=F(u)[\beta _{i,j}c_{1}(u)d_{1,1}(u)+\beta _{j^{\prime },i^{\prime
}}b(u)d_{i,j^{\prime }}(u)]\frac{k_{1,N}(u)}{\beta _{1,N}}\qquad
(i<j^{\prime })  \label{c3}
\end{equation}%
and%
\begin{equation}
k_{i,j}(u)=F(u)[\beta _{i,j}c_{2}(u)d_{1,1}(u)+\beta _{j^{\prime },i^{\prime
}}b(u)d_{i,j^{\prime }}(u)]\frac{k_{1,N}(u)}{\beta _{1,N}}\qquad
(i>j^{\prime })  \label{c4}
\end{equation}%
where%
\begin{equation}
F(u)=\frac{b^{2}(u)-a_{1}(u)d_{1,1}(u)}{%
b^{2}(u)d_{1,2}(u)d_{2,1}(u)-c_{1}(u)c_{2}(u)d_{1,1}(u)^{2}}  \label{c5}
\end{equation}%
and all non diagonal entries $k_{i,j}(u)$ with bosonic degree of freedom ( $%
i,j=m+1,m+2,...,m+r$) in terms of $k_{m+1,N-m}(u)$ 
\begin{equation}
k_{i,j}(u)=B(u)[-\beta _{i,j}c_{1}(u)d_{m+1,m+1}(u)+\beta _{j^{\prime
},i^{\prime }}b(u)d_{i,j^{\prime }}(u)]\frac{k_{1,N}(u)}{\beta _{1,N}}\qquad
(i<j^{\prime })  \label{c6}
\end{equation}%
and%
\begin{equation}
k_{i,j}(u)=B(u)[-\beta _{i,j}c_{2}(u)d_{m+1,m+1}(u)+\beta _{j^{\prime
},i^{\prime }}b(u)d_{i,j^{\prime }}(u)]\frac{k_{m+1,N-m}(u)}{\beta _{m+1,N-m}%
}\qquad (i>j^{\prime })  \label{c7}
\end{equation}%
where%
\begin{equation}
B(u)=\frac{b^{2}(u)-a_{m+1}(u)d_{m+1,m+1}(u)}{%
b^{2}(u)d_{1,2}(u)d_{2,1}(u)-c_{1}(u)c_{2}(u)d_{m+1,m+1}(u)^{2}}.  \label{c8}
\end{equation}%
Note in these expressions that we have used the identity 
\begin{equation}
d_{i,j}(u)d_{j,i}(u)=d_{1,2}(u)d_{2,1}(u),\qquad i\neq j,j^{\prime }.
\label{c9}
\end{equation}%
Simultaneously, we can look at the equations of the type $E[i,i]=0$, with
aid of the property $d_{i,j}(u)=d_{j^{\prime },i^{\prime }}(u)$, in order to
get the symmetric relations 
\begin{equation}
\frac{k_{i,j}(u)}{k_{j,i}(u)}=\frac{\beta _{i,j}}{\beta _{j,i}}%
\Longrightarrow \beta _{i,j}=\beta _{j,i}\frac{\beta _{i^{\prime },j^{\prime
}}}{\beta _{j^{\prime },i^{\prime }}},\quad i>j\quad \mathrm{and\quad }i\neq
j,j^{\prime }.  \label{c10}
\end{equation}%
Next, the equations $E[\acute{\imath},N+1-i]=0$ give us the quadratic
relations ($\epsilon =\pm 1$) between the parameters $\beta _{i,j}$ and $%
\beta _{j^{\prime },i^{\prime }}$: 
\begin{equation}
\beta _{i,j}=-\frac{\epsilon }{\sqrt{\zeta }}\frac{\theta _{i}q^{t_{i}}}{%
\theta _{j^{\prime }}q^{t_{j^{\prime }}}}\beta _{j^{\prime },i^{\prime }}
\label{c11}
\end{equation}%
Substituting these relations into (\ref{RE}), the simplest equation are now
those involving the matrix elements $k_{1,N}(u)$ and $k_{m+1,N-m}(u)$:%
\begin{equation}
\beta _{m+1,N-m}H_{b}(u)k_{1,N}(u)=\beta _{1,N}H_{f}(u)k_{m+1,N-m}(u)
\label{c12}
\end{equation}%
where the $H_{\{f,b\}}(u)$ functions are given by (\ref{Hs}) and we get the
general structure (\ref{Gf}) for the non-diagonal $K$-matrix elements.

Solving each pair of equations $\{E[1,j]=0,E[1,(j-1)N+1]=0\}$ we find the
diagonal matrix elements $\{k_{1,1}(u),k_{j,j}(u)\}$, for $j=2,...,N-1$. The
pair $\{k_{N-1,N-1}(u),k_{N,N}(u)\}$ can be obtained solving the equation $%
\{E[N^{2},N^{2}-1]=0,E[N^{2},N^{2}-N]=0\}$.

\bigskip Again, substituting these relations into (\ref{RE}) we can see that
several equations of the type $E[m+1,j]=0$ give us the constraints between
parameters with bosonic and fermionic entries%
\begin{equation}
\beta _{m+1,b}\beta _{1,f}=0,\quad m+1<b\leq N-m,\quad m+r<f\leq N
\label{c13}
\end{equation}%
It means, in general, that we can get solutions with only a type of degree
of freedom. For a particular choice $\beta _{m+1,b}=0$ or $\beta _{1,f}=0$,
we will find the solutions $\mathcal{M}_{1}$ to $\mathcal{M}_{8}$ presented
in the section $4$ of this paper.

From the reflection equations, we can see for the $U_{q}[osp(1|2m)^{(1)}]$
case that there is no relations of the type (\ref{c1}) with bosonic degree
of freedom and for the $U_{q}[osp(2|2m)^{(1)}]$ case that there is only one
bosonic relation%
\begin{equation}
k_{N-m,m+1}(u)=\frac{\beta _{N-m,m+1}}{\beta _{m+1,N-m}}k_{m+1,N-m}(u)
\label{c14}
\end{equation}%
In addition, for the $U_{q}[osp(r|2)]$ case wee have only one fermionic
relation%
\begin{equation}
k_{N,1}(u)=\frac{\beta _{N,1}}{\beta _{1,N}}k_{N,1}(u)  \label{c15}
\end{equation}%
Therefore, for these cases we can solve the coupled equations (\ref{c12})
without making use of the constraint equations of the type (\ref{c13}).
Thus, we found the four complete solutions $\mathcal{M}_{9}$ to $\mathcal{M}%
_{12}$.

\end{document}